\documentclass[11pt,a4paper,numbers,sort&compress]{article}
\usepackage[utf8]{inputenc}
\pdfoutput=1
\usepackage{jheppub}

\usepackage{physics}
\usepackage{amsmath}
\usepackage{color}
\usepackage{xcolor}
\usepackage{cancel}
\usepackage{amsmath}
\usepackage{amssymb}
\usepackage{slashed}
\usepackage{cancel}
\usepackage[normalem]{ulem}

\usepackage{tikz}

\usepackage{slashed}
\usepackage{cancel}
\usepackage[normalem]{ulem}
\usepackage{multirow}

\usepackage{amsmath, mathrsfs, amssymb,amsfonts,amsthm,graphicx, epsf, dcolumn, yfonts}

\usepackage{subcaption}
\usepackage{graphicx}
\usepackage{color}
\usepackage{slashed}
\usepackage{setspace}
\usepackage{cancel}
\usepackage{wasysym}
\usepackage{float}

\pdfoutput=1
 \newcommand{\be}{\begin{equation}}
 \newcommand{\ee}{\end{equation}}
 \newcommand{\bea}{\begin{eqnarray}}
 \newcommand{\eea}{\end{eqnarray}}

\newcommand{\beq}{\begin{equation}}
\newcommand{\eeq}{\end{equation}}

\renewcommand*{\thefootnote}{\fnsymbol{footnote}}

\begin{document}

\preprint{\texttt{LCTP-24-03, IFT-UAM/CSIC-24-102}}

\title{Fuzzy Spheres in Stringy Matrix Models: \\
Quantifying Chaos in a Mixed Phase Space}

\author{Paolo Amore,$^{1}$ Leopoldo A. Pando Zayas,$^{2}$ Juan F. Pedraza,$^{3}$ Norma Quiroz,$^{4}$ and C\'esar A. Terrero-Escalante$^{1}$}
\affiliation{$^1$Facultad de Ciencias, Universidad de Colima, C.P. 28045 Colima, Colima, Mexico\\
 $^2$Leinweber Center for Theoretical Physics, University of Michigan, Ann Arbor, MI 48109, USA\\
$^2$The Abdus Salam International Centre for Theoretical Physics, 34014 Trieste, Italy\\
$^3$Instituto de F\'isica Te\'orica UAM/CSIC, Calle Nicol\'as Cabrera 13-15, Madrid 28049, Spain\\
$^4$Divisi\'on de Ciencias Exactas, Naturales y Tecnol\'ogicas,
Centro Universitario del Sur, Universidad de Guadalajara
C.P. 49000 Cd Guzman, Jalisco, Mexico}

\abstract{We consider a truncation of the BMN matrix model to a configuration of two fuzzy spheres, described by two coupled non-linear oscillators dependent on the mass parameter $\mu$. The classical phase diagram of the system generically ($\mu \neq 0$) contains three equilibrium points: two centers and a center-saddle; as $\mu \to 0$ the system exhibits a pitchfork bifurcation. We demonstrate that the system is exactly integrable in quadratures for $\mu=0$, while for very large values of $\mu$, it approaches another integrable point characterized by two harmonic oscillators. The classical phase space is mixed, containing both integrable islands and chaotic regions, as evidenced by the classical Lyapunov spectrum. At the quantum level, we explore indicators of early and late time chaos. The eigenvalue spacing is best described by a Brody distribution, which interpolates between Poisson and Wigner distributions; it dovetails, at the quantum level, the classical results and reemphasizes the notion that the quantum system is mixed. We also study the spectral form factor and the quantum Lyapunov exponent, as defined by out-of-time-ordered correlators. These two indicators of quantum chaos exhibit weak correlations with the Brody distribution. We speculate that the behavior of the system as  $\mu \to 0$  dominates the spectral form factor and the quantum Lyapunov exponent, making these indicators of quantum chaos less effective in the context of a mixed phase space.} 

\renewcommand*{\thefootnote}{\arabic{footnote}}
\setcounter{footnote}{0}

\maketitle


\newpage 

\section{Introduction and Summary}

Chaotic phenomena are ubiquitous in both classical dynamical systems and quantum systems. Recently, significant connections between chaos, quantum information, and the quantum aspects of black holes have emerged \cite{Turiaci:2018zsj}. A key example of this synergy is the bound on chaos defined by the quantum Lyapunov exponent \cite{Maldacena:2015waa}. This bound offers a way to compare black holes and quantum many-body systems as information scramblers, highlighting that the saturation of Lyapunov exponents is a common feature of both black holes and certain quantum mechanical systems.

This framework motivates the search for simple quantum mechanical systems that exhibit maximal chaos, offering insights into certain aspects of black hole physics.  This search is bolstered by {\it The Central Dogma}, a viewpoint that has gained prominence through recent advances in our understanding of black holes. As formulated in \cite{Almheiri:2020cfm}, The Central Dogma posits that a black hole, when observed from the outside, can be described by a quantum system of microstates evolving unitarily over time. The search for the simplest realization of this Dogma has occupied part of the community since Kitaev's original proposal in 2015 (see reviews  \cite{Sarosi:2017ykf,Mertens:2022irh}). 

Among the simplest quantum mechanical systems capable of exhibiting black hole-like behavior, one may consider one-dimensional quantum mechanical models. Two notable examples that potentially display black hole-like characteristics in certain limits are the Berenstein-Maldacena-Nastase (BMN) matrix model \cite{Berenstein:2002jq} and the Banks-Fischler-Susskind-Shenker (BFSS) matrix model \cite{Banks:1996vh}. The BFSS model has been demonstrated to encapsulate black hole physics within specific regimes \cite{Hanada:2013rga}, with recent developments and intriguing avenues highlighted in \cite{Maldacena:2023acv}. 

In this manuscript, we study a very drastic truncation of the BMN model. One challenge we encounter is that the resulting system lacks strong chaotic behavior; its phase space is mixed and, therefore, many traditional indicators of chaos yield conflicting results. Consequently, we systematically examine chaos indicators at both classical and quantum levels, focusing on the characterization of weakly chaotic mixed systems. Below, we outline our principal findings.

\subsection{Summary of Results}

\subsection*{Classical Chaos}

After truncating the BMN model, the dynamics is described by the following Lagrangian:
\begin{equation}
\label{Eq:IntroLag}
{\cal L}= \frac{1}{2}\dot{x}^2 + \frac{1}{2}\dot{y}^2 
 -\frac{1}{2}(x^2+y^2)^2 -\frac{1}{8}\mu^2 x^2-\frac{1}{2}\mu^2y^2+ \mu y^3. 
\end{equation}

\begin{itemize}

\item Depending on the value of $\mu$ there are either three (for $\mu >0$) or one  (for $\mu=0$) equilibrium points. The system exhibits a Hamiltonian supercritical pitchfork bifurcation around  $\mu=0$. 

\item We demonstrate that for $\mu=0$ the system is classically integrable by solving it in quadratures. When $\mu$ is much larger compared to other scales in the problem, as indicated by Eq. \eqref{Eq:IntroLag}, the system approximates two decoupled harmonic oscillators.

\item While studying classical chaos, we observed a mixed phase space containing both integrable and chaotic regions, which is a common feature in many dynamical systems.

\item For most regions in phase space we show that there is a positive, albeit small, largest Lyapunov exponent $(\lambda_L \sim  10^{-5})$; we also find initial conditions leading to strongly chaotic trajectories with $\lambda_L \sim 10^{-1}$. 

\item We argue that the nature of such weak chaos is due to a peculiar switching of trajectories between two attractor points, reminiscent of the original Lorenz mechanism, albeit in a system that is Hamiltonian (see Figure \ref{fig:mu5planes} for detailed explanations).
\end{itemize}

\subsection*{Quantum Chaos}

\begin{itemize}

\item We find that the distribution of the energy level spacings, $\Delta E$, is well described by a Brody distribution 

\begin{equation}
    \rho(\Delta E)=c_b\, (\Delta E)^b\exp\left(- c_b\, (\Delta E)^{b+1}\right),
\end{equation}
where $c_b$ is a constant. Note that $b=0$ describes a Poisson distribution while $b=1$ corresponds to a Wigner distribution. The eigenvalue distribution analysis leads to an almost Poison distribution (integrable) $(b\approx 0)$ for $\mu=0$. The value of $b$ then increases as we increase $\mu$, up to $\mu=10$, signaling a more chaotic (Wigner-like) behavior. As we subsequently continue to increase $\mu$ the distribution tends to a Poison-like, signalling integrable behavior. Since the values of $b$ never reach $b=1$ we interpret this fact as an indication of weak quantum chaos.

\item We also study the quantum Lyapunov exponent, following the OTOC prescription. We show that the region of exponential growth of the OTOC is not necessarily well defined and in some cases it simply does not exist. When we are able to reliably compute the quantum Lyapunov exponent this way, we obtain $\lambda_L^{\rm q} \propto T^{0.2}$. This behavior is far from the linear-in-temperature one saturating the bound on chaos. We focus our attention, therefore, on the late-time behavior of the OTOC by performing a spectral analysis. Interestingly, the spectral decomposition shows that the most chaotic configuration is for $\mu=0$ and, as we increase $\mu$, the strength of the indicator decreases.

\item We study the spectral form factor of the model and find that the most chaotic signal in the decay-ramp-plateau paradigm for strongly chaotic systems is displayed for $\mu=0$. As we increase $\mu$ the behavior of the spectral form factor starts resembling the one of harmonic oscillators.

\end{itemize}

A summary of our results is sketched in Table \ref{Table:Summary} which represents the effectiveness of various classical and quantum chaos indicators to detect chaos.

\begin{table}[htbp]
\centering
\begin{tabular}{|c|c|c|p{1cm}p{1cm}p{1cm}|c|}
\hline
 & Chaos Indicator & Pattern & \multicolumn{3}{|c|}{Strength with $\mu$}& Detection  \\ \hline
\multirow{ 2}{*}{Classical Chaos} & Largest Lyapunov  & $\nearrow \searrow$ & Int. & Chaotic & Int. & \checkmark \\ \cline{2-7}
& Phase Space  & $\nearrow \searrow$ & Int. & Chaotic & Int. & \checkmark\\ \hline
\multirow{ 3}{*}{Quantum Chaos} & Brody Distribution  & $\nearrow \searrow$ & Int. & Chaotic & Int. & \checkmark \\ \cline{2-7}
& OTOC  & $\searrow$ & Chaotic & Mixed & Int. & $\times$ \\ \cline{2-7}
& Spectral Form Factor  & $\searrow$ & Chaotic & Mixed & Int. & $\times$ \\
\hline
\end{tabular}
\caption{Summary of various chaos indicators at the classical and quantum level for the BMN matrix model truncated to two fuzzy spheres.}
\label{Table:Summary}
\end{table}

The rest of the manuscript is organized as follows. In section \ref{Sec:Models}, we introduce the model and the particular truncation of interest. In section \ref{Sec:Classical}, we discuss classical aspects of this system and establish that it is weakly chaotic with a mixed phase space. We start the discussion of quantum aspects of chaos in section \ref{Sec:QuantumEigenvalues} with a study of the eigenvalue spacing distribution. Modern indicators of chaos such as the quantum Lyapunov exponent and the spectral form factor are discussed in sections \ref{Sec:OTOC} and  \ref{Sec:SFF}, respectively. We conclude in section \ref{Sec:Conclusions}. We discuss aspects of our numerical method in appendix \ref{app:numerics}.

\section{The BMN Matrix Model Truncated to Two Fuzzy Spheres}\label{Sec:Models}
Let us review the BMN matrix model \cite{Berenstein:2002jq}  highlighting the salient features that will be relevant for our analysis. The BMN matrix model is a one-dimensional $U(N)$ gauge theory composed of matrix-valued variables $X^r(t)$ with $r = 1, \ldots, 9$, a gauge field $A$ and 16 fermions, $\Psi$;  the action of its bosonic sector takes the form:

\be
S=\int dt\,\, {\rm Tr}\bigg[ \frac{1}{2}(D_tX^r)^2 +\frac{1}{4}[X^r,X^s]^2 -\frac{1}{2}\left(\frac{\mu}{3}\right)^2X_i^2
-\frac{1}{2}\left(\frac{\mu}{6}\right)^2X_a^2 -\frac{\mu}{3}i \epsilon_{ijk}X^iX^jX^k\bigg],
\ee
where $r,s=1,\ldots, 9;\, i,j=1,2 ,3$ and $a,b,c=4, \ldots, 9$. The covariant derivative is 
\be
D_t X^r=\frac{d}{dt}X^r -i [A,X^r],
\ee
where $A$ is a $U(N)$ gauge field. 

The equations of motion take the following form:
\begin{eqnarray}
\label{EoM}
\ddot{X}^i&=&\sum\limits_{r=1}^9[[X^r,X^i],X^r]-\left(\frac{\mu}{3}\right)^2 X^i-i\mu \sum\limits_{j,k=1}^3\epsilon^{ijk}X_jX_k, \nonumber \\
\ddot{X}^a&=&  \sum\limits_{r=1}^9[[X^r,X^a],X^r]-\left(\frac{\mu}{6} \right)^2X^a.
\end{eqnarray}
The Gauss law constraint (coming from the equation of the gauge field $A$) implies:

\be
0=\sum\limits_{r=1}^9 [X^r, \dot{X}^r].
\ee 
The fermionic sector will not play a significant role in our analysis. There are supersymmetric configurations in this model that play an important role. Let us review the fully supersymmetric solutions of this action. Imposing that  the supersymmetric variation of the fermion vanishes, $ \delta \Psi = 0$, one  finds that the only solutions are   \cite{Berenstein:2002jq}:
\be
\label{SUSY-solution}
[X^i, X^j]= i\,\, \frac{\mu}{3}  \,\,\epsilon_{ijk}\,\,X^k, \quad i,j,k=1,2,3, \quad  \dot{X}^i=0, \quad X^a=0, \quad a=4,...,9.
\ee

These solutions are known as `fuzzy spheres.' The reason is that for these matrices 
\be
{\rm Tr}\sum\limits_{i=1}^3 (X^i)^2\sim \frac{\mu}{3} N,
\ee
which resembles the equation of a sphere. However, since the `coordinates' are matrices and are inherently non-commutative, the term `fuzzy sphere' is used. The `radius' on the right-hand side above is related to the Casimir of the representation. More precisely, the solutions are labeled by all possible ways of dividing an $N$-dimensional representation of $SU(2)$ into irreducible representations. Note that the fuzzy sphere configuration above becomes trivial in the  $\mu\to 0$ limit. An insightful study of various perturbations around supersymmetric solutions was presented in \cite{Maldacena:2002rb}.

The BFSS model can be regarded as the BMN model in the $\mu \to 0$ limit. The BFSS model is known to suffer from instabilities due to flat directions. For $\mu \neq 0$, however, there are no flat directions because each field is effectively massive. In the $\mu \to 0$ limit, the emergence of flat directions has been suggested to be related to black hole evaporation, as discussed in \cite{Berkowitz:2016znt, Berkowitz:2016muc}. The $\mu \to 0$ limit will play a significant role in this work.

There have been numerous studies on classical chaos in matrix models, starting with Yang-Mills motivated models in \cite{Matinyan:1981dj} and more recently in the context of the BFSS model \cite{Arefeva:1997oyf}. The first exploration of classical chaos within the gauge/gravity correspondence was presented in \cite{PandoZayas:2010xpn}, with subsequent follow-ups and extensions in \cite{Basu:2011di, Basu:2011fw}. Of particular relevance to our investigations is the classical analysis of the BMN model given in \cite{Asano:2015eha}, which presented various aspects of the classical dynamics and established the presence of chaos. Ideally, we would like to connect properties  of chaos in the BMN matrix model with aspects of thermalization. For example, how chaos underlies the role of certain off-diagonal fluctuations \cite{Berenstein:2010bi} and evidence for fast thermalization \cite{Asplund:2011qj} in the BMN matrix model.

We now consider two pulsating fuzzy spheres in the BMN model. Namely, we consider the following Ansatz for the matrices $X^r$  in terms of  $2\times 2$ matrices \cite{Asano:2015eha}:

\bea
X^i&=& y(t)\, \frac{\sigma^i}{2}, \quad (i=1,2,3), \quad X^7=X^8=X^9=0, \nonumber \\
X^{a'}&=& x(t) \frac{\sigma^{a'-3}}{2}, \quad (a'=4,5,6), \nonumber
\eea
where $\sigma^i, i=1,2,3$ are the Pauli matrices.This configuration bears some resemblance to the supersymmetric fuzzy spheres described in \eqref{SUSY-solution}. However, there are crucial differences: (i) our configuration breaks supersymmetry due to its explicit time dependence, and (ii) the matrices $X^{a'}$ do not vanish identically.

The Lagrangian for the effective dynamical variables $x(t), y(t)$ takes the form:
\bea
{\cal L}&=& \frac{1}{2}\dot{x}^2-\frac{1}{8}\mu^2 x^2 -\frac{1}{2}x^4  - x^2 y^2 \nonumber \\
&+& \frac{1}{2}\dot{y}^2-\frac{1}{2}\mu^2y^2+ \mu y^3 -\frac{1}{2}y^4. \nonumber \\
\label{eq:lagrangian}
\eea
Effectively, this is a system of two coupled nonlinear oscillators. 
\subsection{Remarks on Scaling Relations and Regimes}

It will be instructive to examine various scaling properties of the action in certain limits. Note, for example,  that for  $\mu=0$  the quadratic and cubic terms in the Hamiltonian vanish and we end up with:
\begin{eqnarray}
    E&=& p_x^2 +p_y^2 +x^2 y^2 + \frac{1}{2}x^4+\frac{1}{2}y^4.
\end{eqnarray}
The resulting system admits the following scaling transformation:
\begin{equation}
    (x,y)\mapsto (\ell x, \ell y), \quad  t\mapsto \ell^{-1} t, \quad  E\mapsto \ell^4 E.
\end{equation}
Such scaling, as argued in \cite{Akutagawa:2020qbj}, implies that the Lyapunov exponent scales as $\lambda_{classical}\propto E^{1/4}$ since it has dimensions of inverse time. It was further argued that this classical scaling is approximately seen in the quantum treatment of the thermal OTOC. In the next section, we will demonstrate that the $\mu=0$ system is, in fact, classically integrable with important implications for the nature of classical and quantum chaos.

It is worth remarking that the general problem has the following scaling property: 

\begin{equation}
(x,y)\mapsto (\ell x, \ell y), \quad \mu \mapsto \ell \mu, \quad  t\mapsto \ell^{-1} t \quad  \Longrightarrow  \quad E\mapsto \ell^4 E
\end{equation}
Namely, under the re-scaling of the coordinates and parameters above, the energy gets re-scaled in a similar manner. 
One could use this scaling symmetry whenever $\mu\neq 0$ to set $\mu\equiv 1$ and work with the appropriate dimensionless quantities: 
$\left(\frac{x}{\mu}, \frac{y}{\mu}, \mu\, t, \frac{E}{\mu^4}\right)$. 
We will not necessarily follow this approach here but it provides a useful intuition: there are two important cases $\mu=0$ and $\mu\neq 0$; the former one is integrable and the latter one can be studied universally as a function of the effective energy,  $\frac{E}{\mu^4}$.

The regime of large values of $\mu$ is also interesting. Note that when $\mu$ is the largest scale we roughly have two harmonic oscillators around $(x=0,y=0)$. This regime breaks down when $x,y \sim \mu$ in which case the terms $\mu^2 x^2, \mu y^3$ and $x^4,y^4$ become of the same order. 

Therefore, for very small and very large values of $\mu$ the system is nearly integrable. In the next sections, we are going to provide a systematic analysis of the dynamical system given by \eqref{eq:lagrangian}.

\section{Classical Chaos: Mixed Phase Space}\label{Sec:Classical}

The Hamiltonian corresponding to Lagrangian (\ref{eq:lagrangian}) is:
\begin{equation}
H= \frac{1}{2}p_x^2+\frac{1}{2}p_y^2 + \frac{1}{8}\mu^2 x^2 +\frac{1}{2}x^4+x^2y^2 +\frac{1}{2}\mu^2 y^2 -\mu y^3+\frac{1}{2}y^4 \, ,
\label{eq:H}
\end{equation}
and the corresponding Hamiltonian equations of motion are: 
\begin{eqnarray}
\dot{x}&=&p_x\, , \\
\dot{y}&=&p_y\, , \\
\dot{p}_x&=&-\frac{1}{4}\mu^2x-2x^3-2xy^2\, , \\
\dot{p}_y&=&-\mu^2y+3\mu y^2-2x^2y-2y^3\, .
\label{eqs:system}
\end{eqnarray}
The equilibrium points are determined by the vanishing of the equations of motion above. There are either three or one equilibrium points, for  $\mu\neq 0$ or $\mu=0$, respectively. All of these equilibria are located at zero momenta $(p_x=0, p_y=0)$ and $x=0$. Note that the conditions $x=0$ and $y={\rm const.}$, precisely agree with the supersymmetry conditions described in equation \eqref{SUSY-solution}. Note also that the size of the supersymmetric fuzzy sphere is given by the equilibrium value of $y$. The appearance of supersymmetry at the equilibrium points of our system precisely checks our intuition regarding stability of dynamical systems admitting supersymmetry.

More precisely, the system exhibits a Hamiltonian supercritical pitchfork bifurcation. For $\mu=0$ there is a unique equilibrium at $y=0$. If we increase $\mu$ beyond zero,
there appear three equilibria; one located at $y=0$ 
(with eigenvalues $\lambda_{1,2}=\pm i\mu/2$ and $\lambda_{3,4}=\pm i\mu$), 
a center at $y=\mu$ 
(with eigenvalues $\lambda_{1,2}=\pm i\mu/2$ and $\lambda_{3,4}=\pm i\mu$), 
and also a center-saddle at $y=\mu/2$
(with eigenvalues $\lambda_{1,2}=\pm \sqrt{2}\mu/2$ and $\lambda_{3,4}=\pm i\sqrt{3}\mu$/2).

\subsubsection*{Integrability of the $\mu=0$ case}
\label{ssec:m0}

\begin{figure}[!ht]
	\centering
	\includegraphics[width=.5\linewidth]{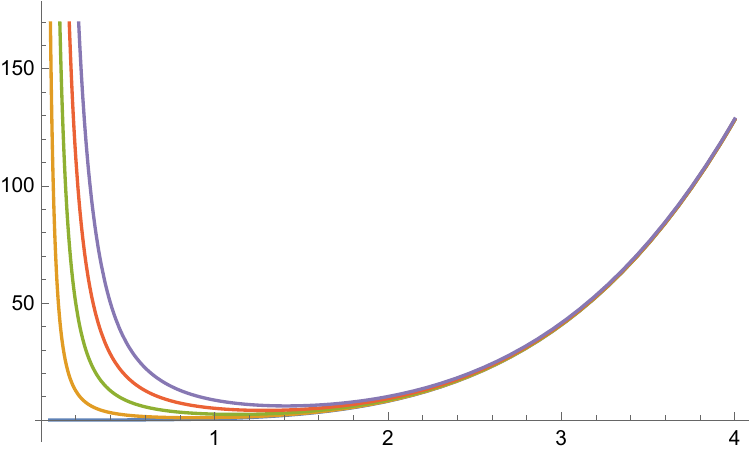}
	\caption{Plot of the effective potential for $\mu=0$ and values of the angular momentum $J=0,1,2,3,4$ (from bottom to top).}
	\label{fig:mu0Pot}
\end{figure}

Since for $\mu=0$ all the eingenvalues of the dynamical system vanish, 
linearizing the vector field around the origin does not help in understanding the dynamics near the equilibrium points. 
Fortunately, {\bf the $\mu=0$ case happens to be integrable}. 
In order to verify this,
let us start by writing the Lagrangian (\ref{eq:lagrangian})  for  $\mu=0$ in polar coordinates:
\begin{eqnarray}
    {\cal L}&=& \frac{1}{2}\dot{x}^2 +  \frac{1}{2}\dot{y}^2-\frac{1}{2}(x^2+y^2)^2 \nonumber \\
&=& \frac{1}{2}\left(\dot{\rho}^2+ \rho^2 \dot{\theta}^2 \right)-\frac{1}{2}\rho^4\, . \nonumber \\
\end{eqnarray}
Note that the Lagrangian does not depend explicitly on time, neither on the coordinate $\theta$, implying that the energy and angular momentum are conserved quantities:
\begin{eqnarray}
    E&=& \frac{1}{2}\dot{\rho}^2 +\frac{J^2}{2\rho^2}+\frac{1}{2}\rho^4 \, , \nonumber \\
    J&=&\rho^2\dot{\theta}\, ,
\label{eqs:mu0cq}    
\end{eqnarray}
with the natural effective potential
\begin{equation}
        V_{eff}\equiv \frac{J^2}{\rho^2}+\rho^4 \, ,
\label{eq:mu0Veff}        
\end{equation}
which is plotted in Figure \ref{fig:mu0Pot}
for different angular momenta.
Therefore, this case can be integrated,
with the general solution given by quadrature,
\begin{equation}
        t = \int \frac{d\rho}{\sqrt{2E-V_{eff}}} \, .
\label{eq:quad}        
\end{equation}

\begin{figure}[!ht]
	\centering
	\includegraphics[width=.5\linewidth]{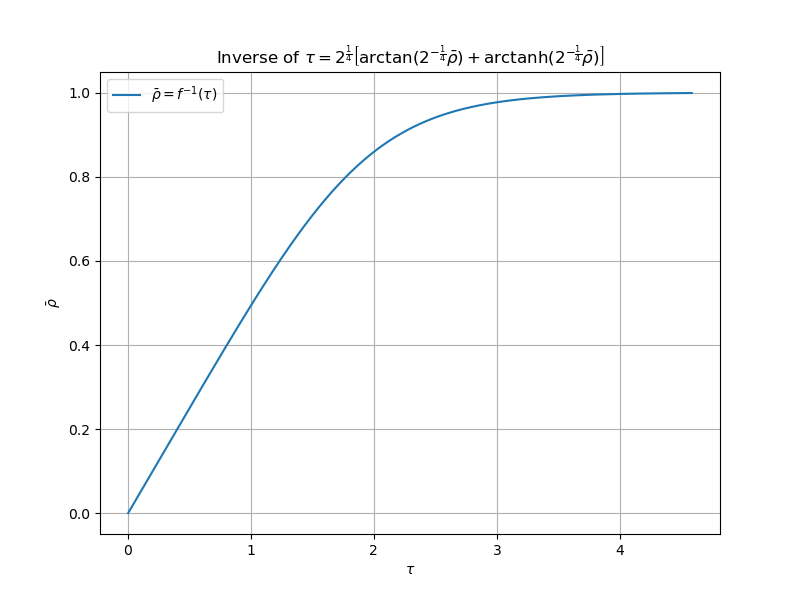}
	\caption{Solution for $\mu=0$ and angular momentum $J=0$.}
	\label{fig:radial}
\end{figure}

\begin{figure}[!ht]
	\centering
\begin{subfigure}{.5\textwidth}
  \centering
  \includegraphics[width=.8\linewidth]{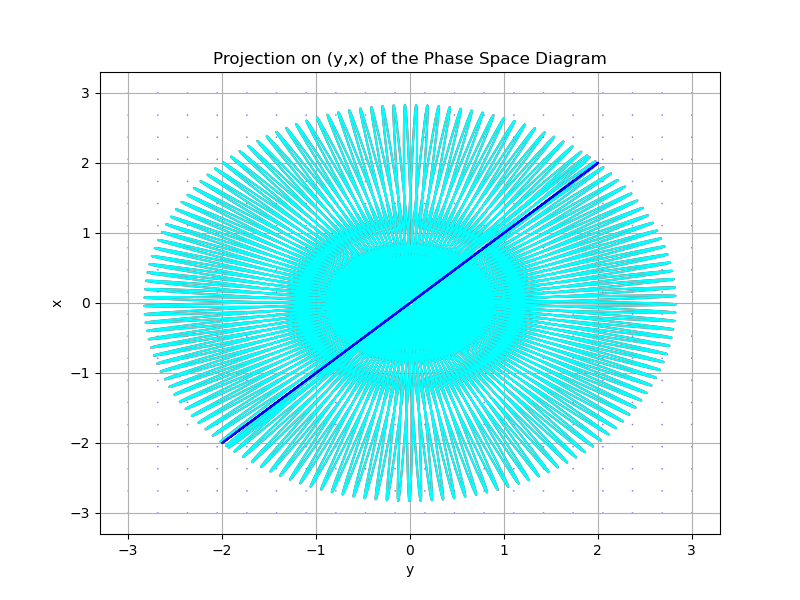}
  \caption{Projection on $(x,y)$.}
  \label{fig:xyplane}
\end{subfigure}%
\begin{subfigure}{.5\textwidth}
  \centering
  \includegraphics[width=.8\linewidth]{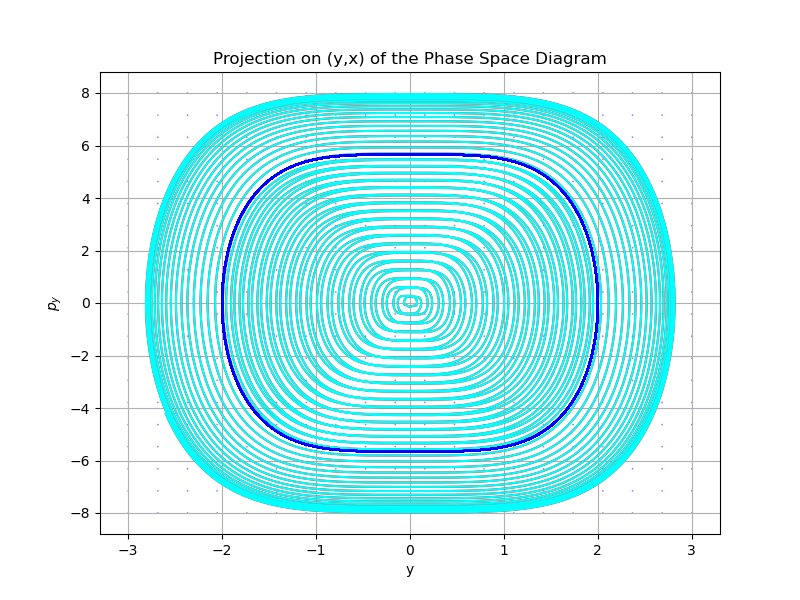}
  \caption{Projection on $(y,p_y)$}
  \label{fig:ypyplane}
\end{subfigure}
	\caption{Projections of the phase-space for $\mu=0$. 
 The dark curve corresponds to $J=0$. 
 Two other trajectories with finite angular momentum are included,
 but only one of them is visible in each figure.}
	\label{fig:mu0planes}
\end{figure}

Note now that for zero angular momentum 
from expressions (\ref{eqs:mu0cq}) and (\ref{eq:mu0Veff})
we get the ordinary differential equation,
\begin{equation}
        \frac{d\bar{\rho}}{d\tau} = \pm \sqrt{1-\bar{\rho}^4}\, ,
\label{eq:rhodot}        
\end{equation}
where we denoted $\tau\equiv (2E)^{1/4} \, t$ and $\bar{\rho} \equiv \rho/(2E)^{1/4}$.
For a positive (negative) sign of the square root
and the real-valued $\bar{\rho}\in [0,1)$,
this equation has a stable (unstable) fixed point at $\bar{\rho}=1$.
Note also that $\bar{\rho}$ can reach zero value,
but with a finite derivative.
For this case the exact solution of (\ref{eq:quad})
is given by 
\begin{equation}
        \tau = F(\arcsin(\bar{\rho})|-1)\, ,
\end{equation}
where $F(\phi|m)$ is the elliptic integral of the first kind. The corresponding behavior may be better understood by noting
that, 
since $\bar{\rho}<1$,
the integrand in \eqref{eq:quad} can be reliably approximated 
by $1/(1-0.5\bar{\rho}^4)$
and the corresponding solution is written as
\begin{equation}
        \tau = f(\bar{\rho}) = 2^{\frac{1}{4}} 
        \left [ \arctan \left ( 2^{-\frac{1}{4}} \bar{\rho} \right ) 
        + {\rm arctanh}\, \left ( 2^{-\frac{1}{4}} \bar{\rho} \right ) \right ]\, .
\end{equation}
The graph of the inverse of function $f(\bar{\rho})$ is presented in Fig.\ref{fig:radial}.
A concrete example of
the corresponding dynamics for variables $(x,y)$ is shown as the 
dark curve in the phase-space projections shown in Figure \ref{fig:mu0planes}.
So, 
for this case, 
trajectories with zero angular momentum 
behave like if the equilibrium at the origin 
of the phase-space
was a center,
with orbits in the $(x,y)$-plane
bouncing radially 
between the two points
with $\bar{\rho}=1$
and periodic orbits lying entirely in a regular two-dimensional torus
in the full four-dimensional phase space.

A slightly more complex behavior arises for finite angular momentum.
As it can be discerned from Figures \ref{fig:mu0Pot} and \ref{fig:mu0planes},
trajectories are now elliptical orbits precessing about the origin.  
In the full 4-dimensional phase-space
these periodic orbits lie on a KAM-torus.
It must be noticed that there are three curves plotted with different colors in each one of these figures.
As already mentioned the dark one corresponds to $J=0$,
with initial conditions $(x_i, p_{x_i}, y_i, p_{y_i})$
given by $(2., -0.1, 2, -0.1)$. 
Two more are included with $(2., -0.1, 2, 0.1)$
and $(2.000001, -0.1, 2, 0.1)$,
respectively.
Since from them 
only one is visible,
it means that
no divergence of nearby trajectories is observed.

\subsubsection*{Bifurcation and chaos for $\mu\neq0$}
\label{ssec:mneq0}

\begin{figure}[!ht]
	\centering
	\includegraphics[width=.75\linewidth]{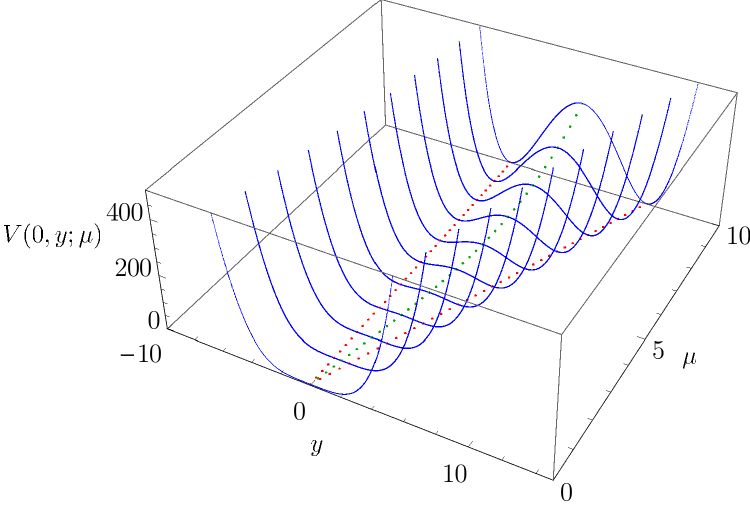}
	\caption{Section of the potential for $x=0$ as a function of $\mu$. Note how the one extremum at $\mu=0$ turns into three extrema as $\mu$ increases.}
	\label{fig:Vvsmu}
\end{figure}

The bifurcation taking place when changing from $\mu=0$ to a finite value 
can be better understood by considering the potential
corresponding to Hamiltonian (\ref{eq:H}),
\begin{equation}
V(x,y;\mu)= \frac{1}{8} \mu^2 x^2 + \frac{1}{2} \mu^2 y^2 - \mu y^3 + \frac{1}{2} x^4 + \frac{1}{2} y^4 + x^2 y^2\, .
\label{eq:V}
\end{equation}
First, it must be noted that it
has a $Z_2$ symmetry with respect to $x$ for any value of $\mu$.
For $\mu=0$ this is also the case in the $y$ direction,
therefore, since the system is Hamiltonian, the solutions correspond to closed iso-level curves around the minimum,
as it was shown in the previous subsection.
Now,
for $\mu\neq0$,
as represented in Fig.\ref{fig:Vvsmu}, 
the $Z_2$ symmetry for $y$ is broken by the $\mu y^3$ term and the behavior of the solutions depends on the initial configuration.

\begin{figure}[!ht]
	\centering
	\includegraphics[width=.75\linewidth]{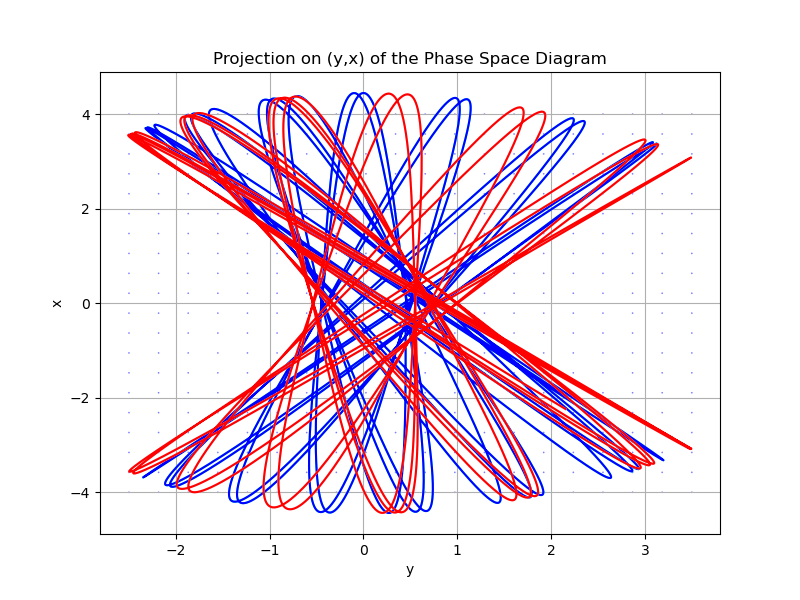}
	\caption{Four solutions for $\mu=1$ and $E\gg V_{max}$.}
	\label{fig:EggVmax}
\end{figure}

If initial conditions
are set such that $E(x_i, p_{x_i}, y_i, p_{y_i}) \gg V_{\rm max}=\mu^4/2^5$,
then very large values of $x_i$ and $y_i$ were used.\footnote{
Another possibility is the total energy
initially
being dominated by the kinetic terms.
However,
because of energy conservation,
in such a case
there will be a time when the corresponding solution 
will reach a point with low kinetic energy
and high potential values.}
It means that the dominant terms in the potential (\ref{eq:V}) are the quartic ones,
so that the high-energy regime can be understood as a small deviation from the exactly integrable $\mu=0$ case.
This is shown in Figure \ref{fig:EggVmax}
where four solutions for $\mu=1$ are presented
with initial conditions $(x_i, p_{x_i}, y_i, p_{y_i})$
 given by $(0., -20, 0.45\mu, 0.)$, $(0.000001, -20, 0.45\mu, 0.)$,
 $(0., -20, 0.55\mu, 0)$, $(0.000001, -20, 0.55\mu, 0)$.
For all these cases $V_{\rm max}=0.03125$.
For the solutions with $x_i=0$, $E=200.030628125$,
while for the ones with $x_i=0.000001$, $E=200.0306281250003$.
For each solution a different color was used in this plot,
but only two curves are visible, 
each one starting at a different side of the maximum.
This means that solutions with very close initial conditions
here are practically indistinguishable. 
The $E\gg V_{\rm max}$ regime emphasizes that $\mu$ is the smallest scale in the problem and the system behaves qualitatively similar to a small deformation of the $\mu=0$ system. 

\begin{figure}[!ht]
	\centering
	\includegraphics[width=.75\linewidth]{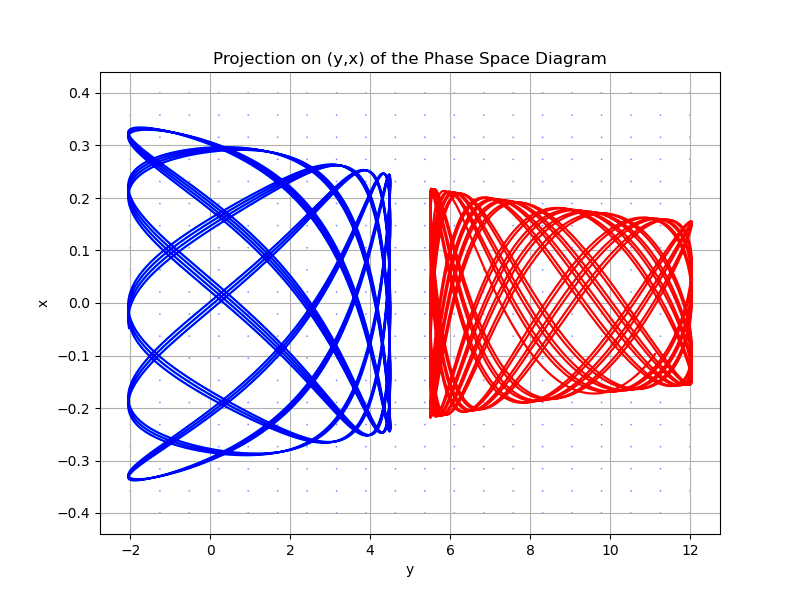}
	\caption{Four solutions for $\mu=10$ and $E< V_{max}$.}
	\label{fig:EllVmax}
\end{figure}

Conversely, the case $E < V_{\rm max}$ can be intuitively understood as the regime of large $\mu$, 
that is, 
when $\mu$ is the largest scale in the problem. 
This regime is approximately integrable and the system is well described by two slightly perturbed harmonic oscillators, as can be inferred from the effective potential \eqref{eq:V}.
In other words,
the solutions will follow periodic orbits
in the full four-dimensional phase space
around the corresponding equilibria,
{\it i.e.},
either around the minimum at $(0,0)$
or around the minimum at $(0,\mu)$.
This is shown in Figure \ref{fig:EllVmax}
where four solutions for $\mu=10$ are presented
with initial conditions $(x_i, p_{x_i}, y_i, p_{y_i})$
 given by $(0., -2, 0.45\mu, 0.)$, $(0.000001, -2, 0.45\mu, 0.)$,
 $(0., -2, 0.55\mu, 0)$, $(0.000001, -2, 0.55\mu, 0)$.
For all these cases $V_{max}=312.5$.
For the solutions with $x_i=0$, $E=308.28125$,
while for the ones with $x_i=0.000001$, $E=308.28125000003$.
Again,
though four different colors were used,
only one curve is visible at each side of the maximum,
implying that solutions with very close initial conditions
remain practically indistinguishable.

\begin{figure}[!ht]
	\centering
\begin{subfigure}{.5\textwidth}
  \centering
  \includegraphics[width=.8\linewidth]{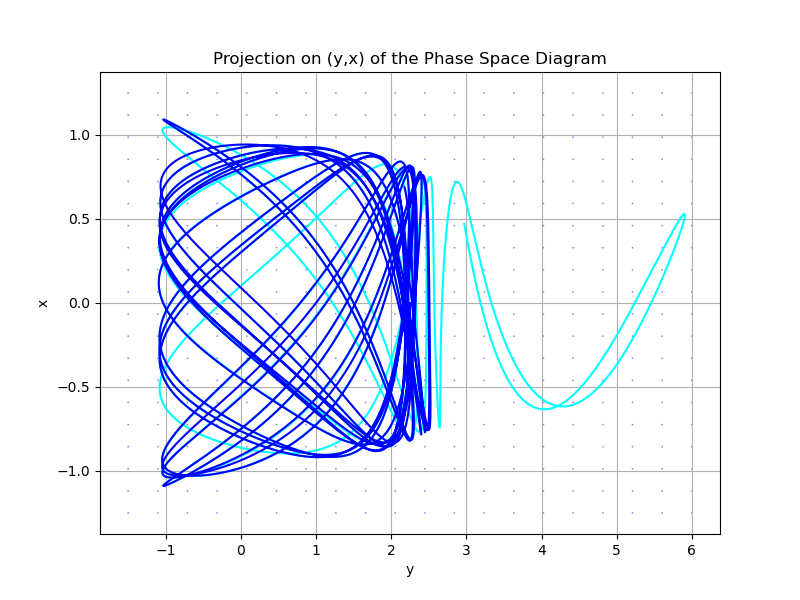}
  \caption{Projection on $(y,x)$ after $t=37$ units.}
  \label{fig:yxplanet40}
\end{subfigure}%
\begin{subfigure}{.5\textwidth}
  \centering
  \includegraphics[width=.8\linewidth]{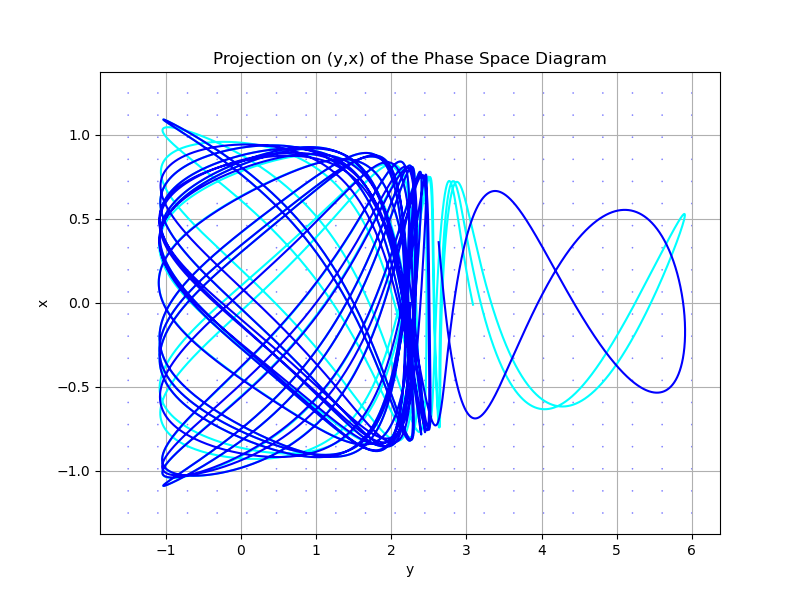}
  \caption{Projection on $(y,x)$ after $t=44$ units.}
  \label{fig:yxplanet161}
\end{subfigure}
	\caption{Projections of the phase-space for $\mu=5$. 
 Two solutions with very close initial conditions are presented at different times.
 }
	\label{fig:mu5planes}
\end{figure}

Now, 
if $E(x_i, p_{x_i}, y_i, p_{y_i}) \approx V_{max}$ from below,
then two orbits starting very close to each other 
may slowly drift apart because the initial separation
will increase while getting closer to the saddle point corresponding to $V_{max}$.  
This explains the chaotic behavior arising in this system
when $E \approx V_{max}$ from above.
If starting to the left of the maximum at $(0,0.5\mu)$,
orbits initially precess around the minimum at $(0,0)$,
eventually reaching the maximum with non-zero momenta,
then they cross to the neighborhood of the minimum at $(0,\mu)$,
circling it until eventually crossing the maximum back into the 
neighborhood of the minimum at the origin.
The interesting feature here is that the number of cycles a solution will do around each minimum
(the time it will take to cross over the maximum)
is very sensitive to the initial conditions
because
orbit instability due to the presence of the saddle point
is now exponentially amplified by the irregular crossing 
from left to right, and viceversa.
As an example, in Figure \ref{fig:mu5planes}
we present two solutions for $\mu=5$
with initial conditions $(x_i, p_{x_i}, y_i, p_{y_i})$
 given by $(0., -3.45, 0.45\mu, 0.)$, $(0.000001, -3.45, 0.45\mu, 0.)$,
 which corresponds to $V_{max}=19.53125$.
For the solution with $x_i=0$, $E=25.093828125$,
while for that with $x_i=0.000001$, $E=25.093828125008$.
As it is shown in the left panel,
after $37$ time units the solution represented by the dark curve
remains circling around the origin,
while the one represented by the light curve
has crossed over the maximum once.
After the light curve bounces back to the left,
both trajectories,
which started very close to each other,
keep circling the origin,
but now following distinctly different paths.
Later on, 
after $44$ time units the solution represented by the dark curve
also crosses to the right of the maximum.
This behavior continues on and renders the distance between the trajectories ever larger. 
Obviously,
the same behavior will be observed if the initial conditions are set to the right of $V_{max}$. 

According to the discussion above, 
we expect for solutions of
system (\ref{eqs:system}) to exhibit mixed behavior, 
with the
ratio of regular to chaotic volumes of the phase space
(which we will judiciously denote as $b$) 
determined by the value of $\mu$.
For very low $\mu$ most of the orbits will be cycling high above $V_{max}$,
while for very large $\mu$ many of the orbits will be cycling around each potential well.
Therefore, it seems like there exists a critical value of $\mu$ for which the phase-space reaches maximal disorder,
i.e., there is a maximum for $b$.

\subsection{Classical Lyapunov Spectrum}

In classical mechanics,
a quantity often used to characterize 
the complexity of the phase-space dynamics 
(regularity and chaos) 
is the largest Lyapunov exponent (LLE). 
It is a measure of the average amount of divergence between nearby trajectories, {\it i.e.}, 
a positive LLE indicates sensitive dependence on the initial conditions, 
one of the common properties of chaos.
For a Hamiltonian system,
there is a whole spectrum of Lyapunov exponents
characterizing the deformation of 
a ball defined by a set of initial conditions
as the system evolves determined by 
Hamilton equations.
According to Liouville's theorem, 
the sum of all these exponents must be zero.

In contrast to the classical examples given 
in the site accompanying the book \cite{Sprott_10.5555/993473},
Hamiltonian systems exhibiting mixed behavior
cannot be characterized by a unique LLE.
As explained in the previous subsection,
whether nearby solutions 
of system (\ref{eqs:system}) 
in average diverge
depends on the region of the phase space probed by the initial conditions.
In general,
given a value of $\mu$,
in the high and low energies regimes
initially close orbits will not diverge exponentially,
while for energies $E(x_i, p_{x_i}, y_i, p_{y_i}) \approx V_{\rm max}$ from above
they will exhibit sensitive dependence on initial conditions.
Several examples of the LLE and Lyapunov spectrum for different initial conditions and values of $\mu$
are given in Table \ref{tab:table01}.
\begin{table}[htp]
	\centering
	\begin{tabular}{||c|l|c|c|c|c|}
		\hline
	& 	$V_{\rm max}=0$ & $x$ & $p_x$ & $y$ & $p_y$ \\
		\hline
 \multirow{ 3}{*}{$\mu=0$}& Initial conditions & 0.1 & 0.06 & 0.0 & 0.4430 \\
  	&	$E = 0.09997$ & & & &\\ 
	&	LLE & 0.00005  & & &\\
	&	Lyapunov spectrum & 0.00012 & 9.9e-5 & -9.2e-5 & -0.00013 \\
		\hline
	\hline
	& 	$V_{\rm max}=0.5$ & $x$ & $p_x$ & $y$ & $p_y$ \\
		\hline
	 \multirow{ 6}{*}{$\mu=2$}	&Initial conditions & -0.2 & 0.1 & 0.0 & 0.4289 \\
   	&	$E = 0.11778$ & & & &\\ 
        &   LLE & 0.00007 & & &\\
	&   Lyapunov spectrum & 0.00015 & 9.3e-5 & -8.5e-5 & -0.00016 \\ 
        &   $\cdots$ & & & &\\
            &Initial conditions & -0.2 & -1.0 & 0.0 & 0.1842 \\
  	&	$E = 0.53776$ & & & &\\   
        &   LLE & 0.06540  & & &\\
	&   Lyapunov spectrum & 0.07110 & 0.00013 & -0.00013 & -0.07110 \\
        		\hline
        \hline
	& 	$V_{\rm max}=5000.0$ & $x$ & $p_x$ & $y$ & $p_y$ \\
		\hline
	\multirow{ 6}{*}{$\mu=20$}	&Initial conditions & 0.4 & 0.0 & 0.0 & 12.0 \\
  	&	$E=80.0128$ & & & &\\  
	&	LLE & 0.00063  & & &\\
	&	Lyapunov spectrum & 0.00082 & 6.3e-05 & -8.1e-05 & -0.00081 \\
        &   $\cdots$ & & & &\\
	&	Initial conditions & 0.4 & 75.0 & 0.0 & -75.0 \\
    	&	$E=5633.0128$ & & & &\\ 
        &   LLE & 0.20614  & & &\\
	&   Lyapunov spectrum & 0.17444 & 0.00189 & -0.00190 & -0.17441 \\
		\hline
 \end{tabular}
\caption{Largest Lyapunov exponent (LLE) and Lyapunov spectrum for several values of $\mu$ and different initial conditions.}
\label{tab:table01}

\end{table}
Taking into account
the finite precision 
of the numerical outputs,
the results presented in this table are consistent with the mixed dynamics of the system, 
as described in the previous subsection.

To calculate the LLE we implemented the algorithm detailed by
Sprott in \cite{Sprott_10.5555/993473} and tested it with the examples there provided in the book.
The Lyapunov spectrum was determined using a version of the algorithm proposed in Ref.\cite{Wolf:1985}.
It was tested with the examples provided in Sprott's book \cite{Sprott_10.5555/993473}.
The LLE and the Lyapunov spectrum are quantities defined asymptotically, 
so for obtaining reliable values
we solved Eqs.(\ref{eqs:system}) using an
implementation of the Dormand-Prince scheme which combines methods of third, fifth and eighth orders to adapt the stepsize, 
ensuring the numerical error to be within a given tolerance.
In turn, 
we use energy conservation to set the parameters of the Dormand-Prince method.
Moreover, since asymptotically the largest positive value in the Lyapunov spectrum must converge to the LLE,
both quantities were used to tune the numerical calculations of each other
(though, 
because of the difference between the numerical methods,
one should expect these numbers to be only approximately equal
),
taking also into account that
the sum of the exponents in the spectrum must be zero
(again,
within the bounded precision 
of the numerical calculations).


\section{Quantum Chaos: Eigenvalue Statistics}\label{Sec:QuantumEigenvalues}
Since the Schroedinger equation is linear, it implies that 
the randomness inherent to a classically chaotic system will not be exhibited by the straightforward time-evolution of its quantum counterpart.
Nevertheless, two conjectures due to Berry-Tabor \cite{9ebf5b64-0d99-3c9d-84d1-37145ab01da7} and Bohigas-Giannoni-Schmit \cite{Bohigas:1983er} allowed to take a significant step into the distinction of quantum versions of regular and chaotic classical systems.
More specifically, the properly normalized 
eigenvalue spacings 
(differences between consecutive eigenvalues, $\overline{\Delta E}$),
for a {\bf generic} integrable system
are distributed according to a {\it Poisson distribution},
\begin{equation}
\rho(\overline{\Delta E}) = e^{- \overline{\Delta E}} \, ,
\label{eq:poisson}
\end{equation}
while for a classically chaotic system, the difference of neighboring eigenvalues is expected to obey the same statistics as for eigenvalues of an ensemble of real symmetric random matrices with independent identically distributed entries.
{\bf Generically}, this is the same probability distribution of the spaces between points in the spectra of nuclei of heavy atoms, {\it i.e.}, it follows the {\it Wigner surmise}:
\begin{equation}
\rho(\overline{\Delta E}) = \frac{\pi \overline{\Delta E}}{2}e^{- \frac{-\pi(\overline{\Delta E})^2}{4}} \, .
\label{eq:wigner}
\end{equation}
Energy level statistics of mixed quantum systems whose classical dynamics are partly regular and partly chaotic are harder to describe. The distinctive feature between the two paradigmatic distributions is the nature of the level repulsion for low eigenvalues present in the chaotic case. For mixed systems, it is convenient to consider the following {\it Brody} distribution: 
\begin{equation}\label{eq:brody}
    \rho(\overline{\Delta E})=c_b\, (\overline{\Delta E})^b\exp\left(- c_b\, (\overline{\Delta E})^{b+1}\right),
\end{equation}
where $c_b \equiv \left[\Gamma\left(\frac{b + 2}{b + 1}\right)\right]^{b + 1}$ is a constant \cite{Brody1973,TomazProsen_1998}. 
The distribution (\ref{eq:brody}) interpolates between Poisson ($b=0$) and Wigner ($b=1$) distributions. 
As anticipated in section \ref{Sec:Classical},
the value of $b$ gives us an estimation of the
ratio of regular to chaotic volumes of the phase space. 

For accurate statistical analyses of the distribution of eigenvalue spacings, it is crucial to properly account for symmetries
present in the system. 
Symmetries can create independent sub-spectra with different statistical properties within the full spectrum. 
For the Hamiltonian (\ref{eq:H}), 
$x\to-x$ invariance 
means that we can label the eigenstates as having either even or odd parity
with respect to the $x$-coordinate.
It is also very important
to remove the global trends in the spectrum, 
allowing to focus on local fluctuations.
In our study,
taking this into account is particularly relevant
because results from the classical analysis in Sec.\ref{Sec:Classical}
point to levels with $E_i\approx V_{\rm max}$
as playing a special role,
which can be obscured by the global properties of the spectrum.
Unfolding is a technique 
commonly used 
to ensure that we can compare the local statistical properties of the spectrum 
with theoretical predictions, such as those from Random Matrix Theory.
It is also very important to take into account that
the unfolding procedure is not unique. 
The spacings statistics is very sensitive to the choice 
of the unfolding function in polynomial unfolding.
Therefore,
we tested different methods and settings,
finally choosing a configuration
yielding results in better agreement
with the analysis in Sec.\ref{Sec:Classical}.
Summarizing,
for our study
we first
decouple Hamiltonian (\ref{eq:H}) into matrices corresponding to each symmetry class.
Then we solve the corresponding Schr\"odinger equations
using the numerical method described in details in Appendix \ref{app:numerics}.
Finally,
for each subset of eigenvalues $\{E_i\}$, 
we use a polynomial interpolation
as a smooth approximation of the corresponding spectral staircase function.
This allows to map the original eigenvalues to unfolded eigenvalues.
Then we fit the histogram of the spacings of consecutive $\{E_i\}_{\rm unf}$,
using Huber weights to reduce the influence of the outliers on the fitting process. We test the strength of expression (\ref{eq:brody}) by introducing an additive parameter $\epsilon$ to measure the departure of data statistics 
from Brody distribution. In all of the considered cases we obtained $|\epsilon|<10^{-3}$. Since the largest difference between Poisson and Wigner distributions
arise at the smaller spacings,
where the density is close to $1$,
then any additive correction to (\ref{eq:brody})
can be neglected.

\begin{figure}[!ht]
	\centering
\begin{subfigure}{.5\textwidth}
  \centering
  \includegraphics[width=.8\linewidth]{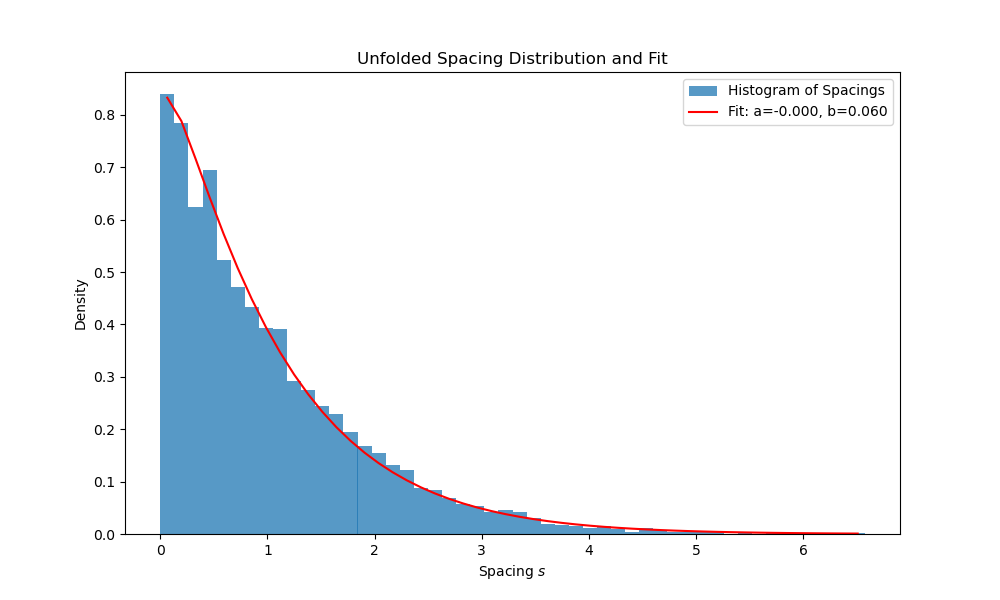}
  \caption{Even modes, $b = 0.06$.}
  \label{fig:brodymu0even}
\end{subfigure}%
\begin{subfigure}{.5\textwidth}
  \centering
  \includegraphics[width=.8\linewidth]{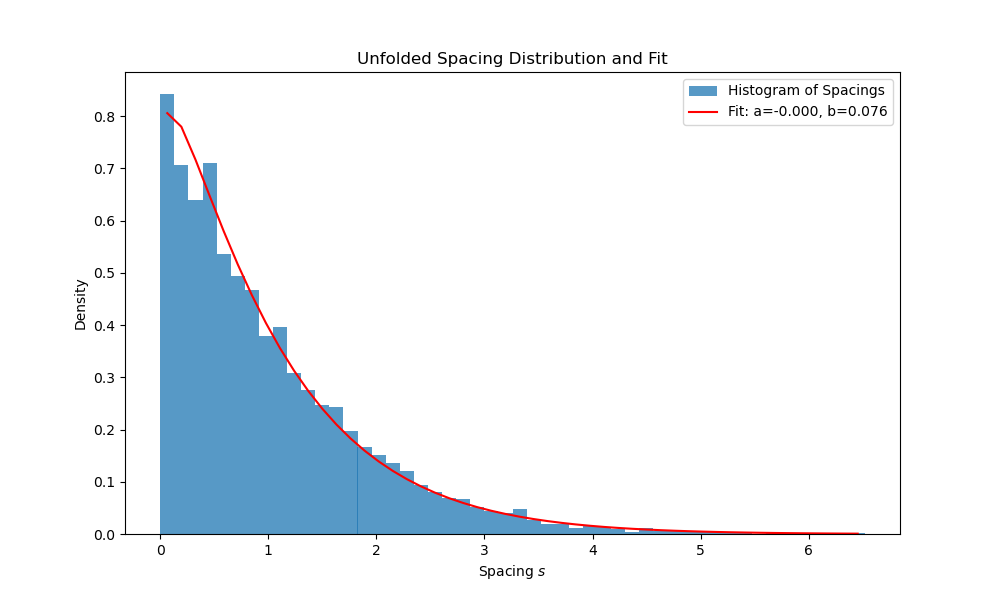}
  \caption{Odd modes, $b = 0.08$.}
  \label{fig:brodymu0odd}
\end{subfigure}
	\caption{Spacings distribution for $\mu=0$.
 }
	\label{fig:brodymu0}
\end{figure}

\begin{figure}[!ht]
	\centering
\begin{subfigure}{.5\textwidth}
  \centering
  \includegraphics[width=.8\linewidth]{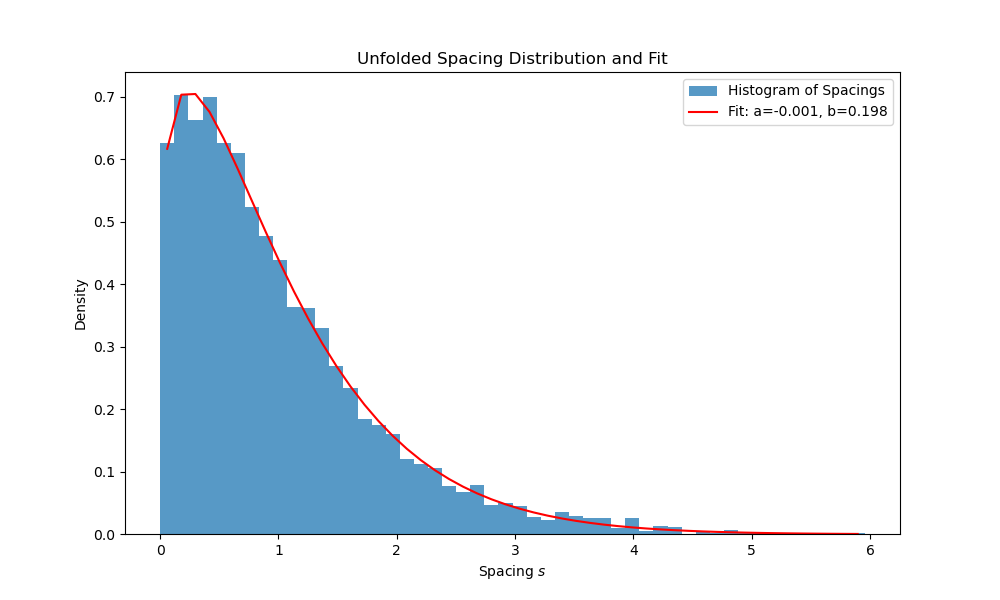}
  \caption{Even modes, $b = 0.20$.}
  \label{fig:brodymu2even}
\end{subfigure}%
\begin{subfigure}{.5\textwidth}
  \centering
  \includegraphics[width=.8\linewidth]{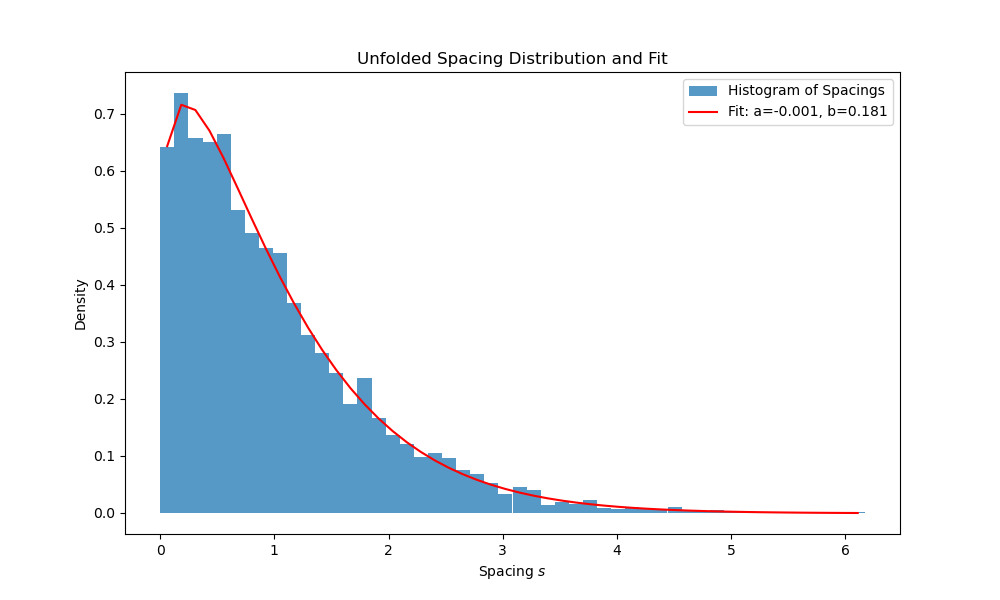}
  \caption{Odd modes, $b = 0.18$.}
  \label{fig:brodymu2odd}
\end{subfigure}
	\caption{Spacings distribution for $\mu=2$.
 }
	\label{fig:brodymu2}
\end{figure}

\begin{figure}[!ht]
	\centering
\begin{subfigure}{.5\textwidth}
  \centering
  \includegraphics[width=.8\linewidth]{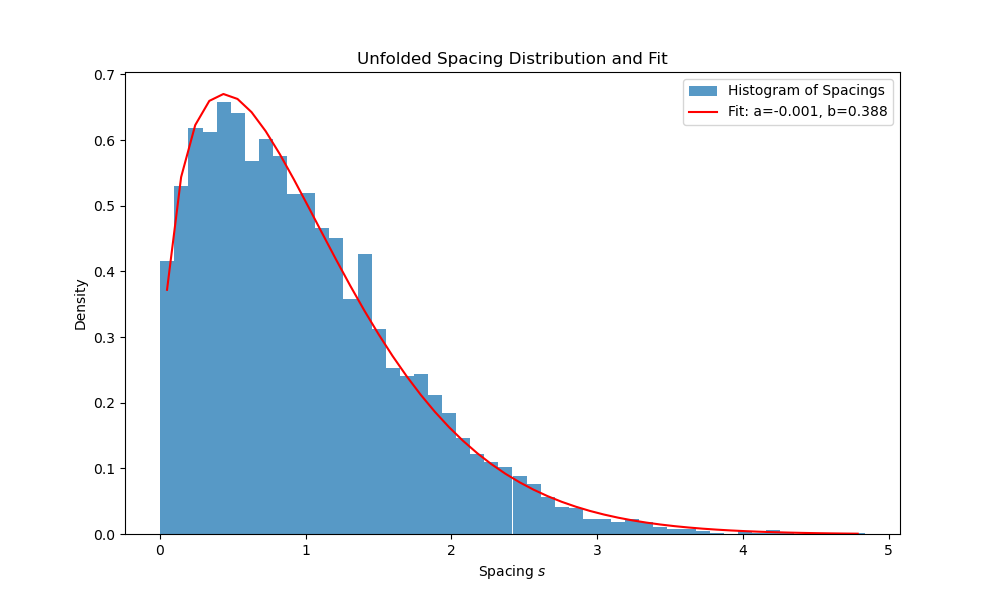}
  \caption{Even modes, $b = 0.39$.}
  \label{fig:brodymu5even}
\end{subfigure}%
\begin{subfigure}{.5\textwidth}
  \centering
  \includegraphics[width=.8\linewidth]{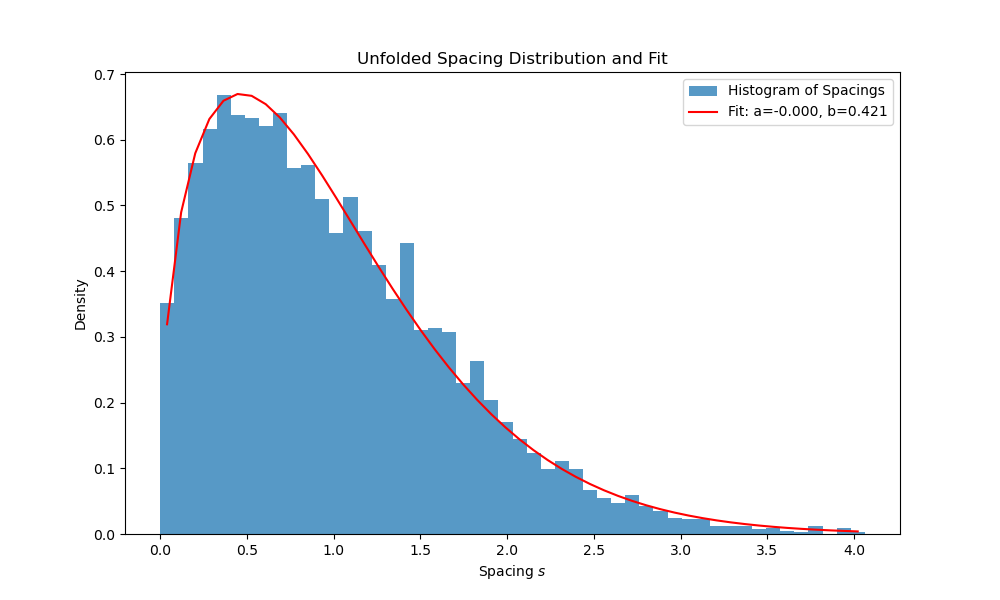}
  \caption{Odd modes, $b = 0.42$.}
  \label{fig:brodymu5odd}
\end{subfigure}
	\caption{Spacings distribution for $\mu=5$.
 }
	\label{fig:brodymu5}
\end{figure}

\begin{figure}[!ht]
	\centering
\begin{subfigure}{.5\textwidth}
  \centering
  \includegraphics[width=.8\linewidth]{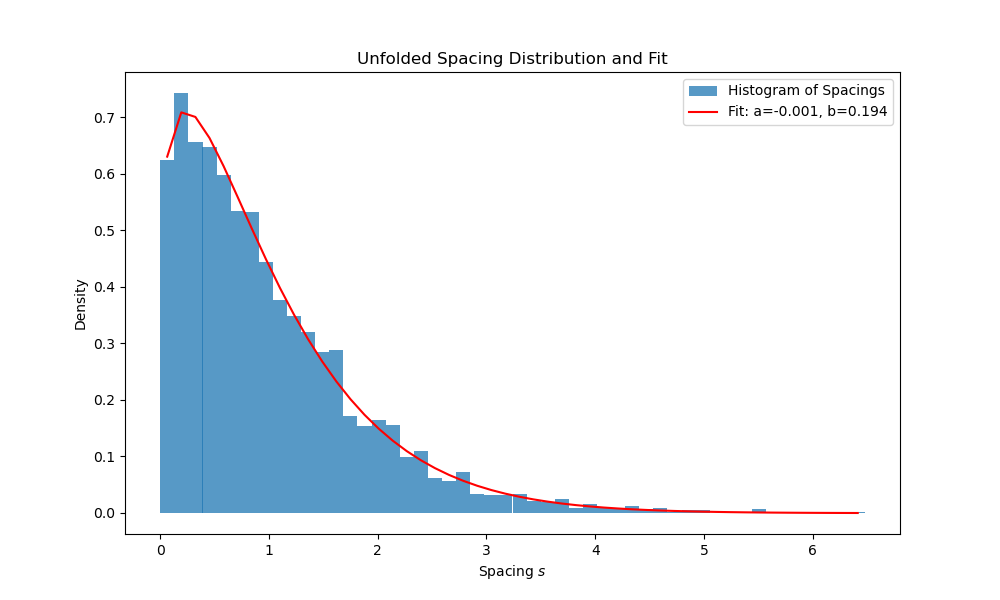}
  \caption{Even modes, $b = 0.19$.}
  \label{fig:brodymu10even}
\end{subfigure}%
\begin{subfigure}{.5\textwidth}
  \centering
  \includegraphics[width=.8\linewidth]{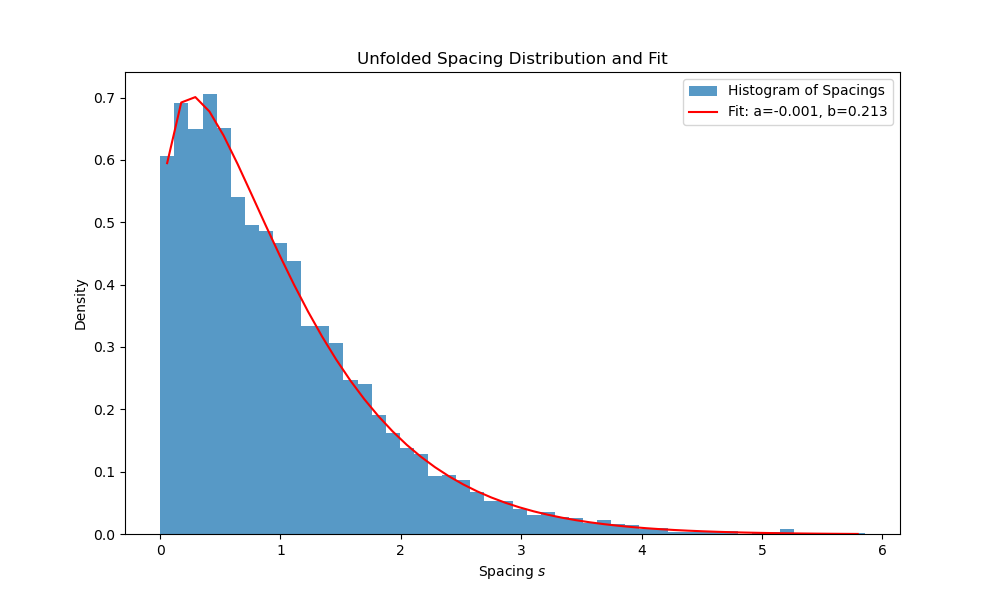}
  \caption{Odd modes, $b = 0.21$.}
  \label{fig:brodymu10odd}
\end{subfigure}
	\caption{Spacings distribution for $\mu=10$.
 }
	\label{fig:brodymu10}
\end{figure}

\begin{figure}[!ht]
	\centering
\begin{subfigure}{.5\textwidth}
  \centering
  \includegraphics[width=.8\linewidth]{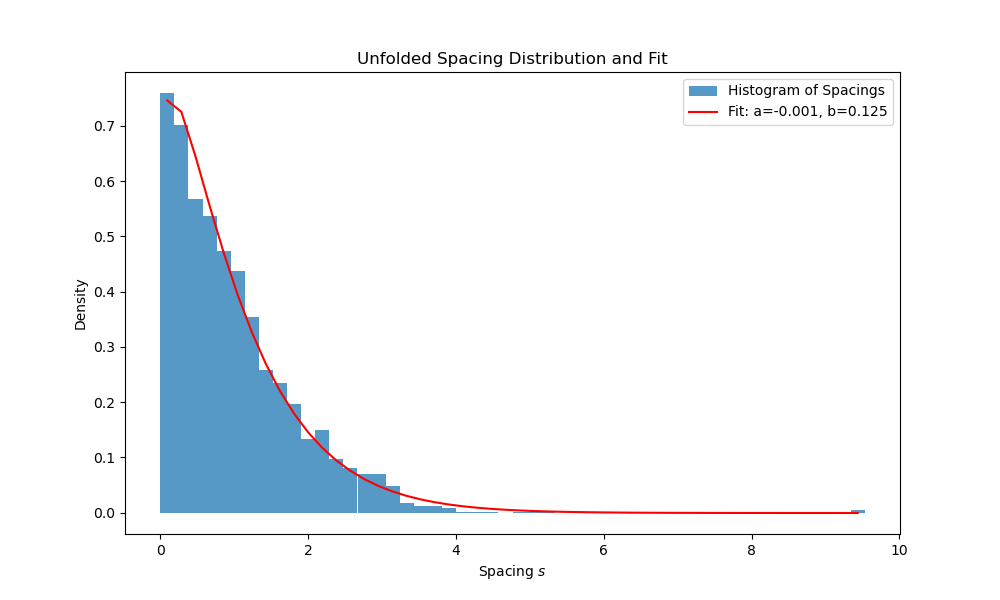}
  \caption{Even modes, $b = 0.13$.}
  \label{fig:brodymu20even}
\end{subfigure}%
\begin{subfigure}{.5\textwidth}
  \centering
  \includegraphics[width=.8\linewidth]{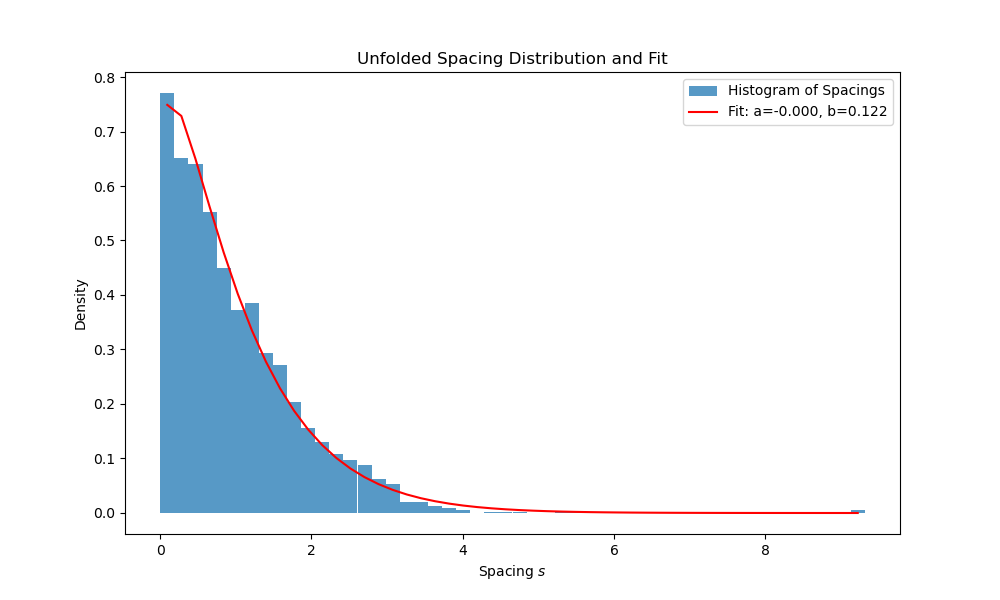}
  \caption{Odd modes, $b = 0.12$.}
  \label{fig:brodymu20odd}
\end{subfigure}
	\caption{Spacings distribution for $\mu=20$.
 }
	\label{fig:brodymu20}
\end{figure}

The obtained best-fit values of the Brody parameter $b$
for some values of $\mu$
are presented in Table \ref{tab:tableBrody}.
\begin{table}
    \centering
    \begin{tabular}{||c|c|c|c|c|c||}
    \hline
        $\mu$  & 0 & 2 & 5 & 10 & 20\\
    \hline        
         b (even) & 0.06 & 0.20 & 0.39 & 0.19 & 0.13 \\
    \hline         
         b (odd) &  0.08 & 0.18 & 0.42 & 0.21 & 0.12 \\         
    \hline         
    \end{tabular}
    \caption{Bestfit values of Brody parameter $b$ as function of $\mu$.}
    \label{tab:tableBrody}
\end{table}
In turn,
the density distribution behavior corresponding to these values
can be observed in Figures 
\ref{fig:brodymu0},
\ref{fig:brodymu2},
\ref{fig:brodymu5},
\ref{fig:brodymu10}
and
\ref{fig:brodymu20}.
For $\mu=0$
the spacings statistics,
for even and odd modes,
closely follows a Poisson density distribution.
This picture starts to slowly change as $\mu$ is increased,
as can be observed, 
for instance,
in Figure \ref{fig:brodymu2},
where the results for $\mu=2$ are shown.
Moving up,
for $\mu=5$ the density distribution
(Fig.\ref{fig:brodymu5})
looks very much like that predicted by Wigner surmise.
However,
as it can be observed in Fig.\ref{fig:brodymu10},
increasing $\mu$ further leads back to distributions with lower $b$.
Finally,
the spacings statistics 
for cases with large $\mu$
are practically indistinguishable
from those with $\mu=0$
(for instance,
$\mu=20$
in Fig.\ref{fig:brodymu20}
).

Within the precision of the numerical calculations,
this behavior of the density distribution 
of the quantum eigenvalues
is in agreement with the classical analysis 
presented in Sec.\ref{Sec:Classical}.
There it was proved that
at $\mu=0$ the system is integrable, 
for small values of $\mu$
the solutions 
can be reasonably approximated by
the $\mu=0$ quadrature,
while for large values of $\mu$ 
the system behaves as two decoupled harmonic oscillators, 
thus approaching another integrable limit.

\section{Quantum Lyapunov Exponents from OTOC}\label{Sec:OTOC}

Let us briefly recall how the quantum Lyapunov exponent is typically introduced. A useful probe of the early development of quantum chaos is motivated by the quantum butterfly effect as described in \cite{Shenker:2013pqa}. Namely, one estimates the strength of such effects by considering the following quantity: 
\be\label{eq:Coft}
C(t)=-\left\langle [W(t),V(0)]^2\right\rangle\,,
 \ee
where $\langle\cdot\rangle\equiv Z^{-1}\text{tr}[e^{-\beta H}\cdot]$ denotes the thermal expectation value and $\beta$ is the inverse temperature. Physically, the commutator squared diagnoses the effect of perturbations by generic operators $V$ on later measurements of $W$. For chaotic quantum systems, $C(t)$ grows exponentially up to a time scale, dubbed `Ehrenfest' or scrambling time $t_{*}$, at which the initial perturbation has spread out within an $\mathcal{O}(1)$ fraction of the system's degrees of freedom. The
quantity $C(t)$ can be expressed as a combination of time-order and out-of-time-order (OTOC) correlators. The former are insensitive to chaos, as they decay at a shorter time scale, the dissipation time $t_d$, such that $\langle V(0)V(0)W(t)W(t)\rangle\sim\langle VV\rangle\langle WW\rangle+\mathcal{O}(e^{-t/t_d})$. The OTOCs, on the other hand, can evolve over longer times. Specifically, the chaotic properties of (\ref{eq:Coft}) can be diagnosed by computing
\be\label{OTOCdef}
f(t)=\frac{\langle V(0)W(t)V(0)W(t)\rangle}{\langle VV\rangle\langle WW\rangle}\,.
\ee
For chaotic systems with a large number of degrees of freedom $N_{\text{dof}}$, there can be a parametrically large hierarchy between the scrambling and dissipation times. This allows for a time window $t_d\ll t\ll t_*$ during which (\ref{OTOCdef}) may exhibit a well-defined exponential evolution, such that
\be\label{Eq:QLyap}
f(t)=1-\frac{f_0}{N_{\text{dof}}}e^{\lambda_L t}+\mathcal{O}(N_{\text{dof}}^{-2})\,.
\ee
Analogous to classical chaos, the exponent $\lambda_L$ in (\ref{Eq:QLyap}) is referred to as the quantum Lyapunov exponent; it obeys a universal bound \cite{Maldacena:2015waa}
\be
\lambda_L\leq \frac{2\pi}{\beta}\,,
\ee
which is saturated for field theories with Einstein gravity duals.\footnote{The bound is saturated in both the closed and open string sectors. In the former case, it can be attributed to the scattering of gravitons in the near-horizon region \cite{Shenker:2014cwa}. In the latter, it results from the scattering of an infinite tower of string excitations, which effectively mimic single graviton exchange \cite{deBoer:2017xdk}.} The dissipation time here is set by the lowest quasinormal mode of the host black hole $t_d\sim\beta$, while $t_*\sim \beta\log N_{\text{dof}}$. 

In quantum mechanical systems with few degrees of freedom, there is generally no significant hierarchy between the scrambling and dissipation times. As a result, depending on other scales of the problem, a regime of exponential evolution may or may not exist, and the quantum Lyapunov exponent may or may not be well-defined. Our goal here is to investigate this issue within the context of the truncated BMN matrix model.

To compute the OTOC in a quantum mechanical system we follow the algorithm outlined in \cite{Hashimoto:2017oit}. For concreteness, and by analogy with the classical case, we pick the operators as $V=\hat{p}$ and $W=\hat{x}$. The calculation requires us to compute the energy spectrum, $E_n$, and the matrix elements, $x_{nm}$, which we obtained numerically, to very high accuracy. Details of these numerical calculations are given in Appendix \ref{app:numerics}. We compute the microcanonical commutator squared $c_n(t)\equiv -\langle n | [\hat x(t),\hat p(0)]^2 |n \rangle$ which can be expressed as: 
\begin{eqnarray}\label{eq:micro_otoc}
c_n(t)&=&\sum_m b_{nm}(t)b^\ast_{nm}(t) \, , \\
 b_{nm}(t)&=&\frac{1}{2}\sum_{k} x_{nk}x_{km} (E_{km}e^{iE_{nk}t}-E_{nk}e^{iE_{km}t})\,, \qquad   E_{nm}\equiv E_n-E_m\,.
\end{eqnarray}
Finally, we compute the thermal expectation value 
\begin{equation}\label{Eq:Micro-to-Thermal}
    C(t)=Z^{-1}\sum_n e^{-\beta E_n} c_n(t)\,.
\end{equation}

\subsection{Microcanonical and Thermal OTOC  }
To calculate the OTOC in the quantum mechanical system of interest, we follow the strategy outlined in \cite{Hashimoto:2017oit} and reviewed earlier. There are several important numerical aspects that need clarification. 

First, let us discuss the criteria used to truncate the spectrum consistently.  To estimate the number of eigenvalues we need to keep to achieve numerically stable results, we utilize the Boltzmann factor, $e^{-E_n/T}$. Note that precisely this factor enters in going from the microcanonical OTOC to the thermal OTOC \eqref{Eq:Micro-to-Thermal}.  Clearly, the number of terms needed depends on the temperature, with higher temperatures requiring more terms for a given precision. Typically, for higher eigenvalues $E_n$, this Boltzmann factor is close to zero. We numerically considered non-vanishing terms with a tolerance of $10^{-3}$.  Due to computational limitations, we focused on temperatures $T \leq 8$ for $\mu=0,2,5$. In these cases, the contributions of the Boltzmann factor to the thermal OTOC come only from $n \leq 200=N_{\rm trunc}$.   For larger values of $\mu=10, \ldots,20$, $N_{\rm trunc}$ is much smaller as the temperature is raised and the numerical requirements are less stringent.

The next step is to introduce a truncation level $M_{\rm trunc}$ in the sum over the subindex $m$ in the microcanonical OTOC \eqref{eq:micro_otoc}. For the selected value of $N_{\rm trunc}$ as described above, we start by plotting the microcanonical OTOC with small values of $M_{\rm trunc}$ and then gradually increase it. The appropriate truncation level for computing the thermal OTOC is identified at the point where the microcanonical curves converge. In our case, $M_{\rm trunc} = 300$ for  $\mu=0,2,5$.  For $\mu = 10,20$ smaller values of $M_{\rm trunc}$ suffice. With these parameters at hand, we then proceed to compute and plot both the microcanonical and thermal OTOC.

\begin{figure}[t]
\centering
\subfloat[]{\includegraphics[width=0.45\textwidth]{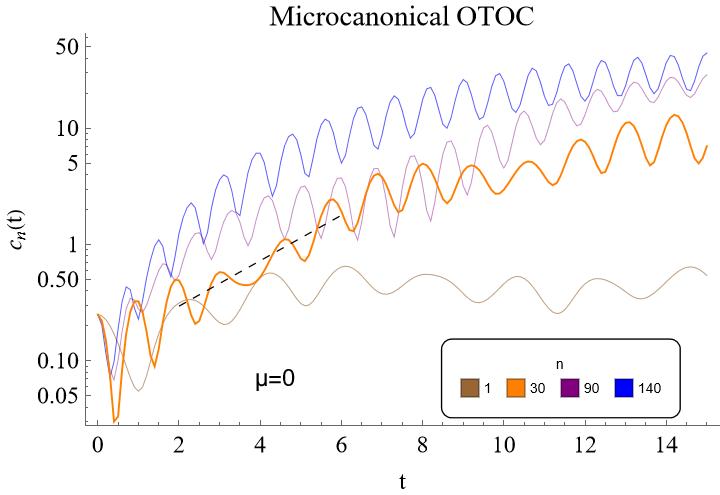} \label{fig:micromu0}} \hfill
 \subfloat[]{\includegraphics[width=0.45\textwidth]{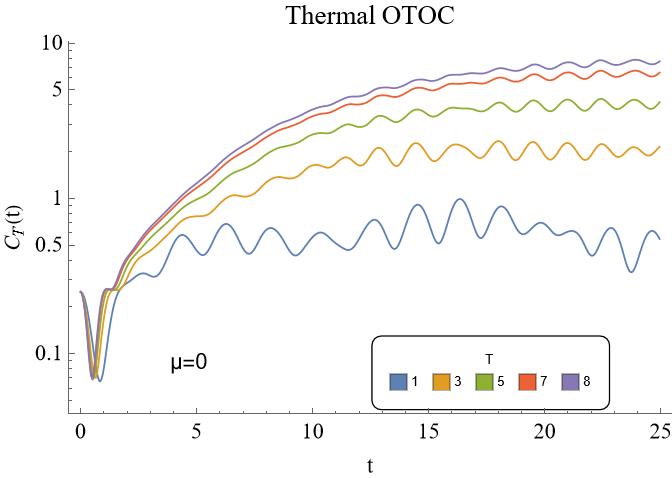}\label{fig:ctimu0}}\\
  \subfloat[]{\includegraphics[width=0.45\textwidth]{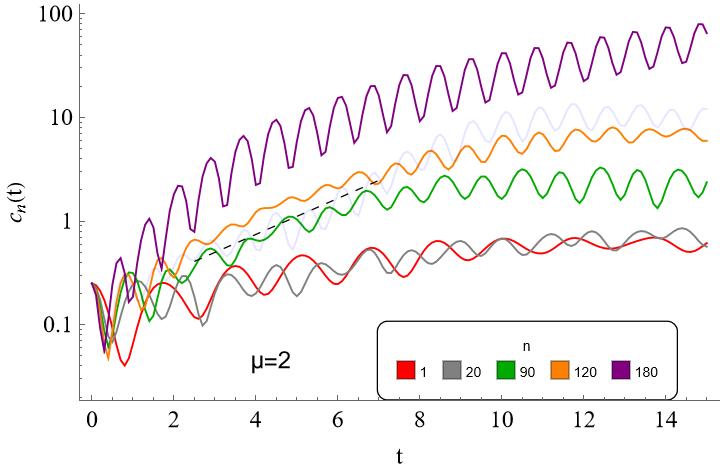}\label{fig:micromu2}}  \hfill
 \subfloat[]{\includegraphics[width=0.45\textwidth]{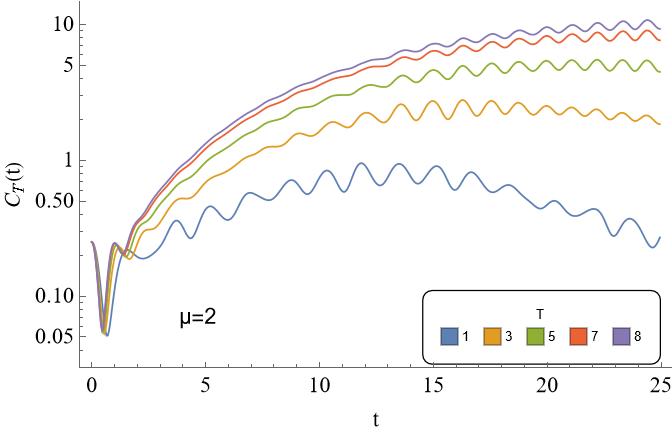}\label{fig:ctimu2}}\\
 \subfloat[]{\includegraphics[width=0.45\textwidth]{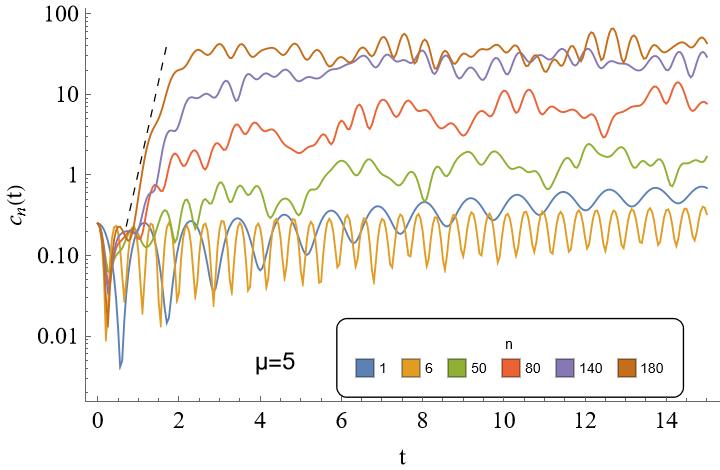}\label{fig:micromu5}}  \hfill
  \subfloat[]{\includegraphics[width=0.45\textwidth]{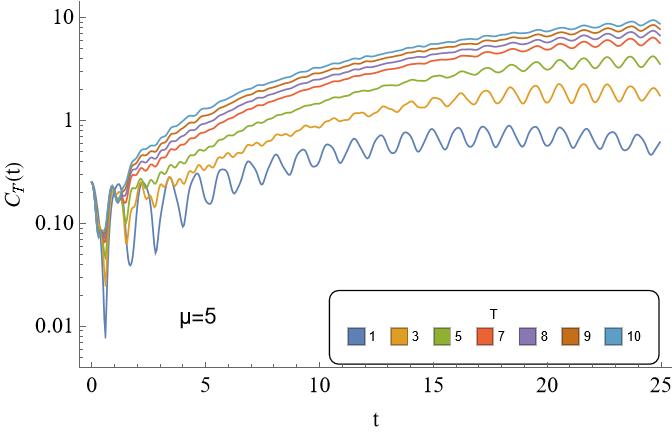}\label{fig:ctimu5}} \\
\caption{Short-time behavior in the logarithmic scale for microcanonical  OTOC (first column)  and thermal OTOC (second column). A weak exponential growth is observed as a scaling region in the microcanonical plots  for some values of $n$. For the thermal OTOC there is no clear linear growth.}
\label{fig:shortCT025}
\end{figure}

\begin{figure}[t]%
  \centering
\subfloat[]{\includegraphics[width=0.45\textwidth]{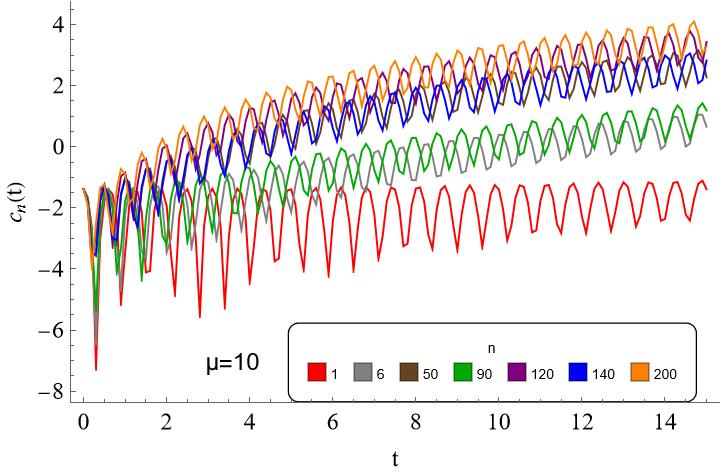}\label{fig:micromu10}}\hfill
 \subfloat[]{\includegraphics[width=0.45\textwidth]{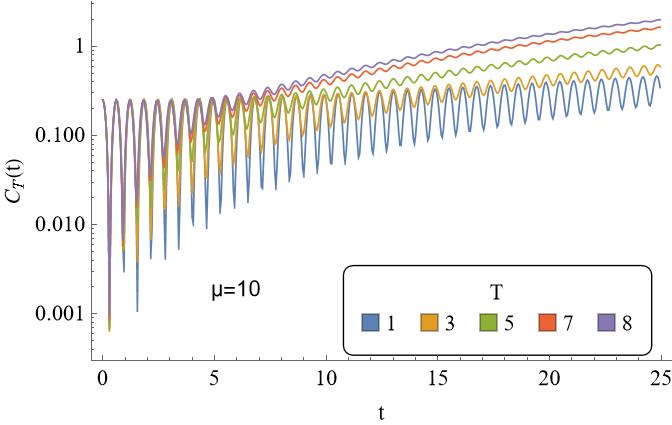}\label{fig:ctimu10}}\\
 \subfloat[]{\includegraphics[width=0.45\textwidth]{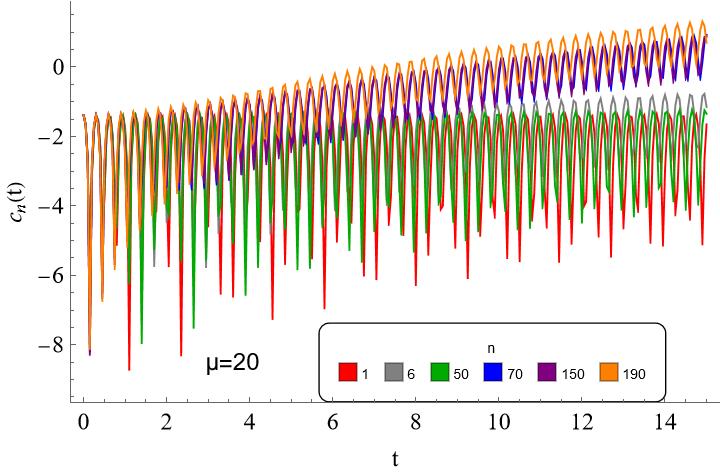}\label{fig:micromu20}}\hfill
 \subfloat[]{\includegraphics[width=0.45\textwidth]{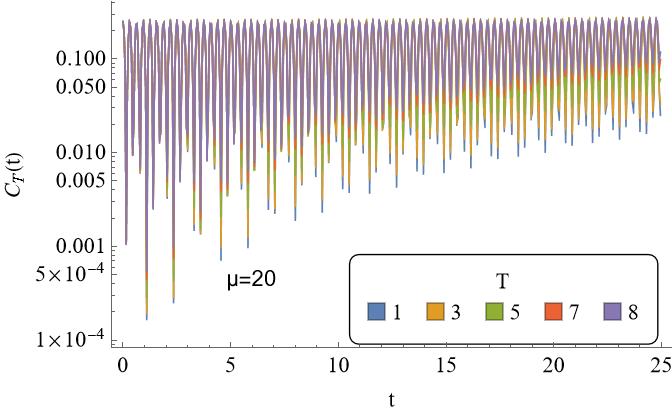}\label{fig:ctimu20}}
 \caption{Short-time behaviour in the logarithmic scale for microcanonical  OTOC (first column)  and thermal OTOC (second column).} 
 \label{fig:shortCT1020}
\end{figure}

Following \eqref{Eq:QLyap}, we then look for an exponential growth in the plot of $C(t)$. Identifying this range of exponential growth in time from our model is one of the main technical challenges we face. The numerical results of our study are displayed in Fig.~\ref{fig:shortCT025} and Fig.~\ref{fig:shortCT1020},  which show the temporal evolution of the microcanonical (left panels) and thermal OTOC (right panels) on a linear-logarithmic scale for $\mu = 0,2,5$ and $\mu=10,20$ respectively. We first explored the microcanical OTOC plotting this function for several values of $n < 200$.  For $\mu=0$ and 2, the behavior of the curves is diverse but 
 for some values of  $n$ a slight linear growth between $t = 2$ and $t=4$ is observed as shown in Fig.~\ref{fig:micromu0} for $n=30$  and in  Fig.~\ref{fig:micromu2}  for $n=90, 120$. On the other hand, for $\mu=5$ there is a distinct scaling region for  several  values of $n$ within the small interval between $t=1$ and $t=2$  as depicted in  Fig.~\ref{fig:micromu5}. For $\mu = 10,20$ the microcanonical OTOC  is predominantly characterized by strong oscillations.

 For the thermal OTOC, we plotted the functions for several values of the temperature and we did not observe a clear exponential growth. Initially, all curves exhibit strong fluctuations between $0 \leq t \leq 2$ which we attribute to dissipation. For $T =1$ $C_T(t)$ does not show significant growth and continue to display strong oscillations. For $T=3$ the best fit behavior in the range $2 \leq t \leq 10$  appears to be a power law. As the temperature is raised,  the curves exhibit a minor step-like increase around  $ 1.5 \leq t \leq 1.7$, with this effect more pronounced for $\mu=5$. It may be inferred that there is a brief period of exponential growth in $C_T(t)$ for $2 \leq t \leq 3.5$ before transitioning to a power law behavior. However, this interval is too short to definitively distinguish between exponential and power law growth.

\begin{figure}[t]%
  \centering
 \includegraphics[width=0.55\textwidth]{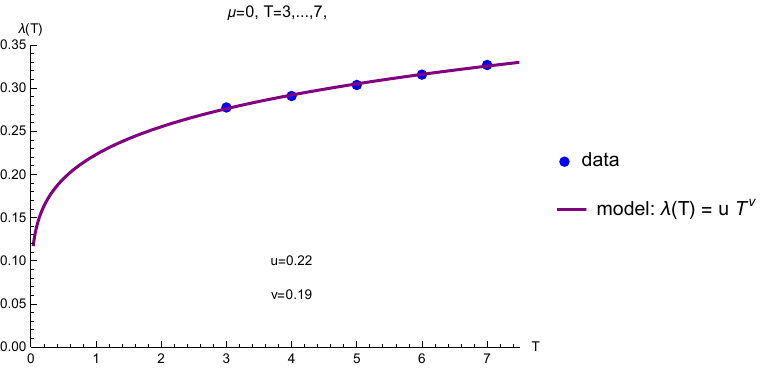}
 \caption{ The Estimation of the Lyapunov exponent $\lambda$ for  $\mu=0$  by fitting the curves to the model   $C_T(t)= a e^{t \lambda(T)}$ between $t=2-3.5$}. 
  \label{fig:lyap}
 \end{figure}

Despite these challenges, we conducted an exponential fit for $\mu=0$ using  Least-Square regression. The Lyapunov exponent was fitted with the power law function $\lambda = u\, T^v$. The behavior for temperatures $T=3,3.5,\ldots,7$ is depicted in Fig.~\ref{fig:lyap} with $u = 0.22$,  $v=0.19$ and a sum of squares of $6.7 \times 10^{-6}$.  
  
\subsection{Late-time Behavior of OTOC}

We have examined the growth phase of the OTOC, yet this appears to be a transient and challenging region to unequivocally characterize. Several studies have noted that for systems lacking maximal chaos, and few-body systems, the late-time behavior of the OTOC may serve as a more suitable indicator of sub-maximal chaos. A systematic approach has been recently proposed in \cite{Garcia-Mata:2018slr,Fortes:2019frf}, with further discussion in \cite{Markovic:2022jta}. For a broader review of OTOCs in quantum chaos, particularly focusing on late-time dynamics, refer to \cite{Garcia-Mata:2022voo}.

\begin{figure}[t]%
\centering
 \subfloat[]{\includegraphics[width=0.4\textwidth]{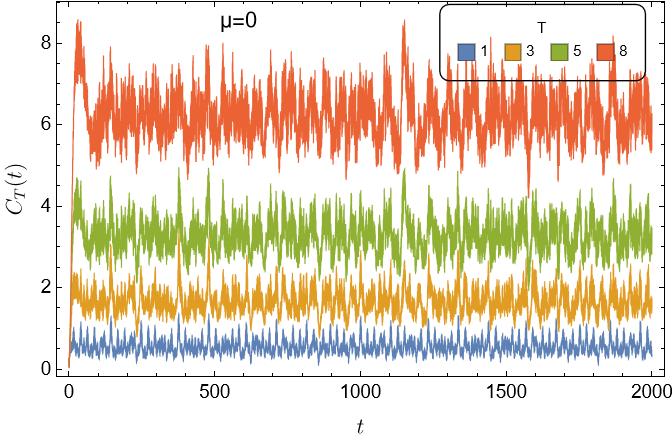}\label{fig:ctmu0}}%
 \hfil
  \subfloat[]{\includegraphics[width=0.4\textwidth]{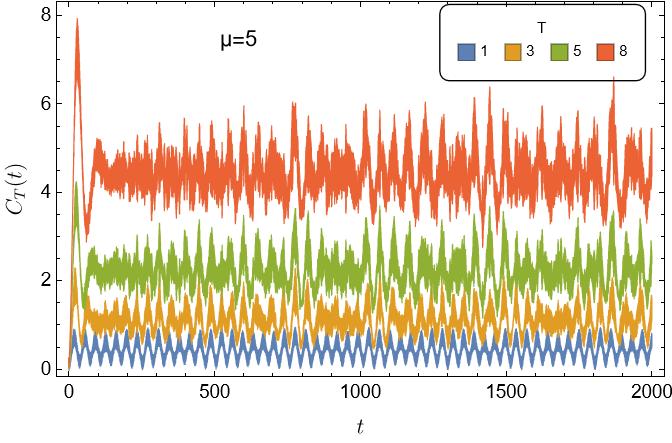}\label{fig:ctmu5}}\\
  \subfloat[]{\includegraphics[width=0.4\textwidth]{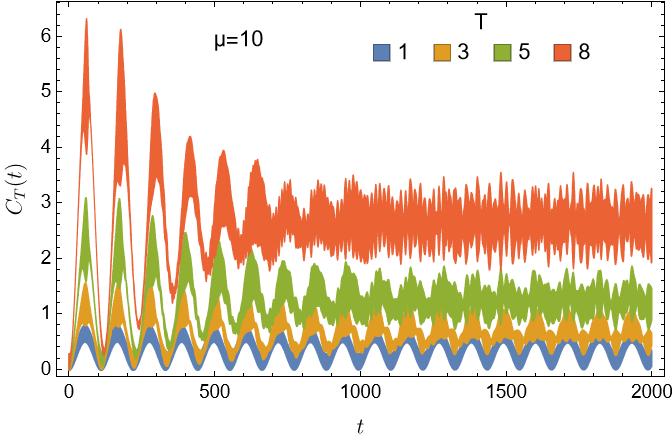}\label{fig:ctmu10}}%
 \hfil
 \subfloat[]{\includegraphics[width=0.4\textwidth]{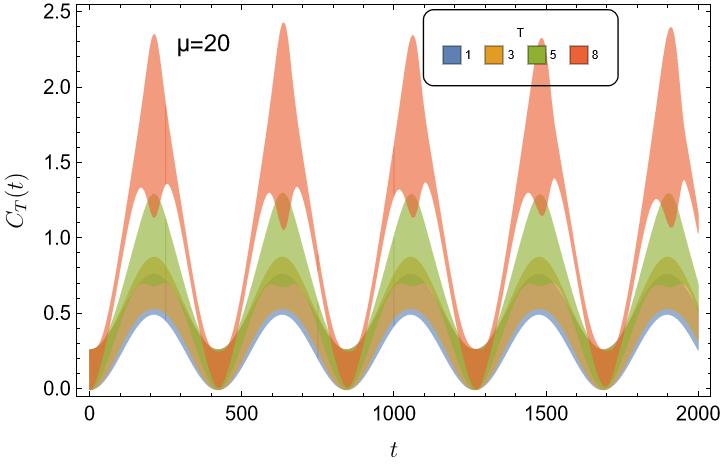}\label{fig:ctmu20}}
 \caption{Late-time plots of the thermal OTOC.}\label{Fig:Late-OTOC}
 \end{figure}

In Fig. \ref{Fig:Late-OTOC} we depict the behavior of several representative OTOCs up to times $t=2000$. The curve for  $\mu=10$ shows quasi-periodic oscillations at early times, roughly below $t=700$. However, it transitions to irregular behavior at later times, approximately $750 \lesssim t \lesssim 2000$. In contrast, for $\mu=20$, the behavior appears to remain periodic throughout.

\begin{figure}[t]%
\centering
\subfloat[]{\includegraphics[width=\textwidth]{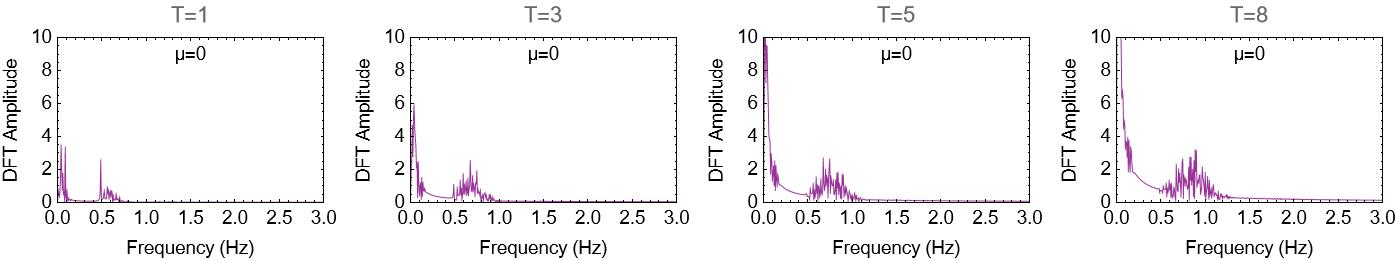}\label{fig:ctmu0}}\\
 \subfloat[]{\includegraphics[width=\textwidth]{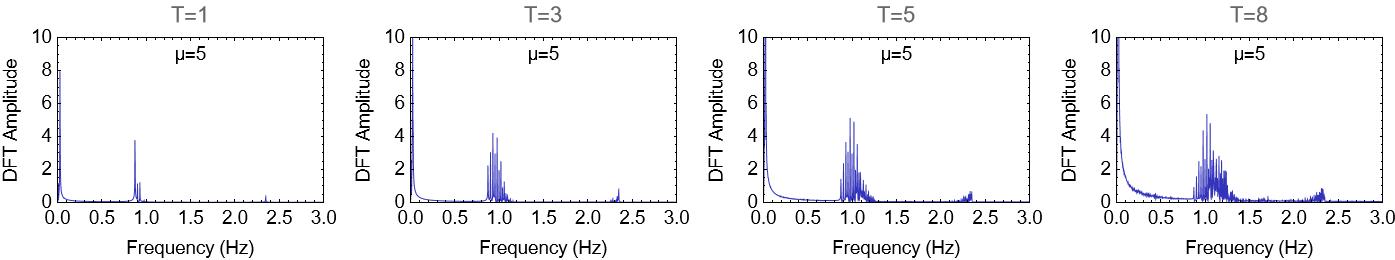}\label{fig:ctmu5}}\\
 \subfloat[]{\includegraphics[width=\textwidth]{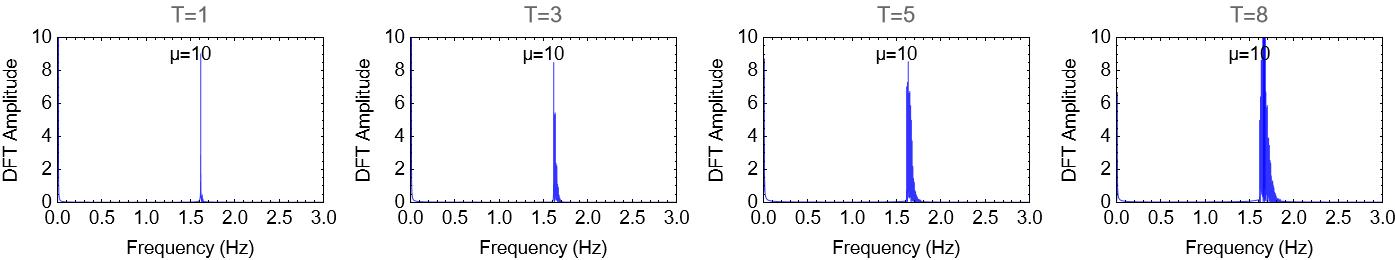}\label{fig:ctmu10}}\\
 \subfloat[]{\includegraphics[width=\textwidth]{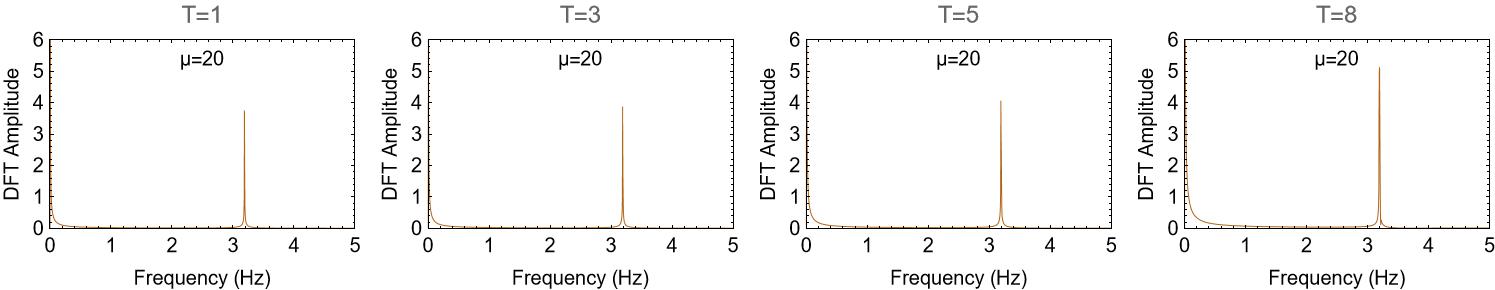}\label{fig:ctmu20}}\\
  \caption{Frequency participation in the late-time OTOC signal} \label{fig:ct}
  \end{figure}

The plots in Fig. \ref{Fig:Late-OTOC} are, of course, inherently difficult to interpret directly. In Fig. \ref{fig:ct} we present the amplitude spectrum corresponding to the late-time data of the thermal OTOC shown previously in Fig. \ref{Fig:Late-OTOC}. This spectrum illustrates the frequency components contributing to the time signal derived from the late-time OTOC. 

From Fig. \ref{fig:ct} it is evident that as we increase the temperature for any value of $\mu$, the number of frequencies contributing to the signal decreases. This can be interpreted straightforwardly: higher temperatures lead to a more monochromatic signal, indicating less chaos. Moreover, as $\mu$ increases from $\mu=0$ to $\mu=20$, Fig.\ref{fig:ct} shows a similar trend where fewer frequencies participate in the signal. Essentially, higher values of  $\mu$ correspond to less chaotic behavior in the late-time OTOC signal.

\section{Spectral Form Factor}\label{Sec:SFF}

As a final probe of quantum chaos, we will explore the behavior of the spectral form factor (SSF), a quantity constructed from the analytically continued partition function. 

The motivation for considering this observable also stems from studies of black holes, particularly from the observation that the behavior of horizon fluctuations in large anti-de Sitter (AdS) black holes mirrors the random matrix dynamics seen in quantum chaotic systems \cite{Cotler:2016fpe}. This study relies on the Sachdev-Ye-Kitaev (SYK) model, a simplified representation of a black hole, whose Hamiltonian is given by
\begin{equation}
H=\frac{1}{4!}\sum_{a,b,c,d}J_{abcd}\psi_a\psi_b\psi_c\psi_d\,.
\end{equation}
Here $\psi_i$ are Majorana fermions and $J$ are random couplings satisfying a Gaussian distribution. By employing numerical methods, \cite{Cotler:2016fpe} precisely characterized the random matrix behavior of the SYK model at early and late times, utilizing the SSF as a diagnostic tool.

The spectral form factor is defined as follows:
\be
g(t)=Z(\beta+it)Z(\beta-it)\,,\label{eq:sff}
\ee
where $Z(\beta)$ is the standard thermal partition function,
\be
Z(\beta)\equiv \sum_i e^{-\beta E_i}\,.\label{eq:tpf}
\ee
Assuming the energy spectrum is chaotic, the SSF 
exhibits a universal slope-ramp-plateau pattern, which is exactly reproduced by the SYK model \cite{Cotler:2016fpe}, as shown in Fig. \ref{fig:SFF} below.

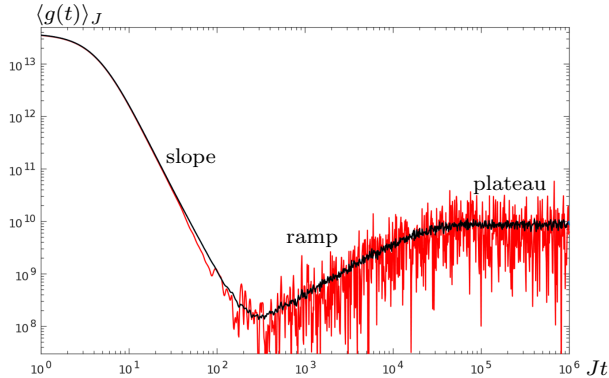
\begin{figure}[h!]
\centering
 \includegraphics[width=3in]{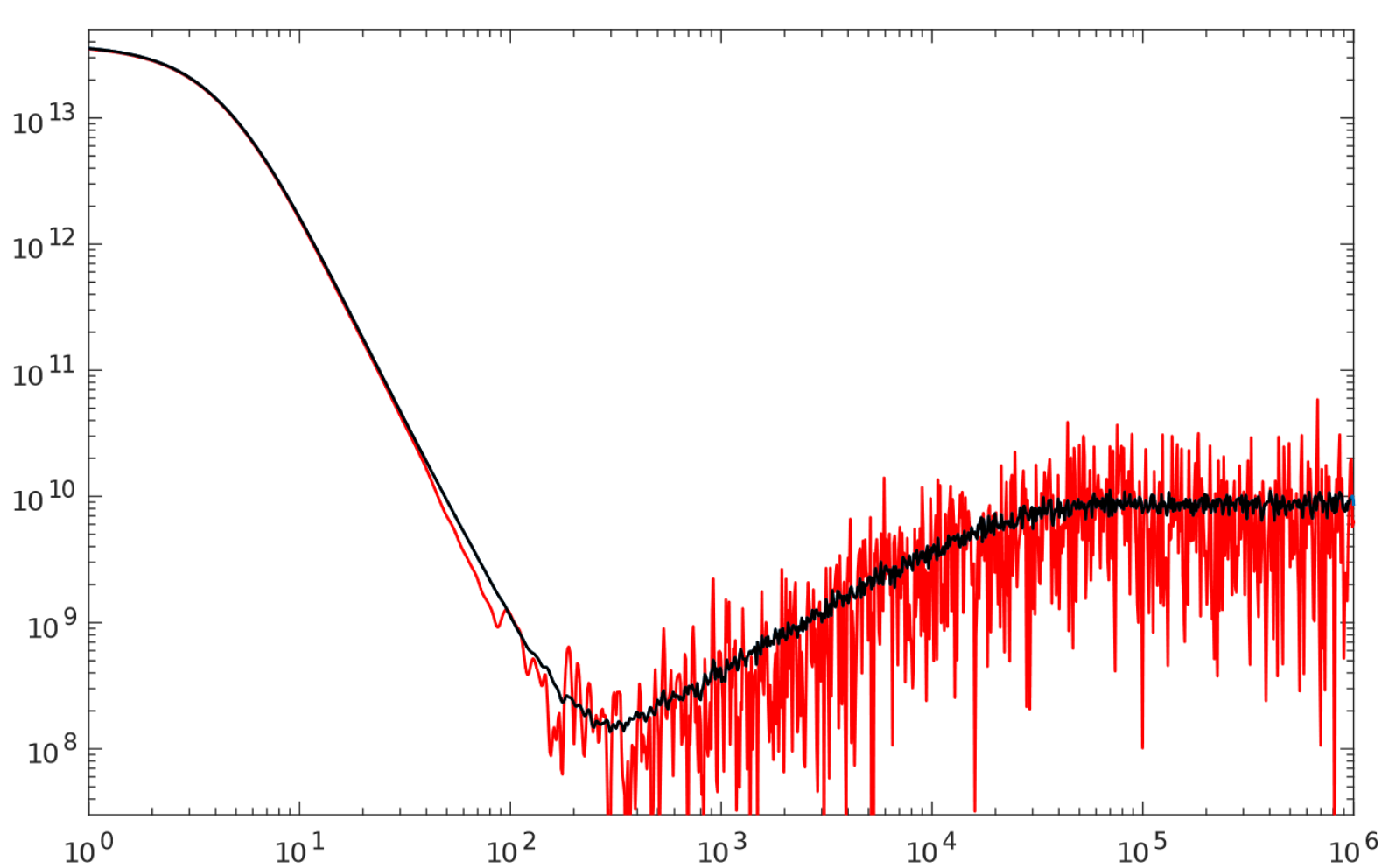}
 \begin{picture}(0,0)
\put(-209,134){\scriptsize$\langle g(t)\rangle_J$}
\put(-3,0){\scriptsize$J t$}
\put(-160,80){\scriptsize slope}
\put(-115,50){\scriptsize ramp}
\put(-45,70){\scriptsize plateau}
\end{picture}
\caption{Typical log-log plot of the spectral form factor in the SYK model, matching the expectation from 2d gravity \cite{Cotler:2016fpe}. The brackets $\langle\cdot\rangle_J$ indicate an ensemble average over the space of couplings. When the average is taken over a large sample, the late-time fluctuations become softer and ultimately disappear (alternatively one can also consider an average over short time scales).\label{fig:SFF}}
\end{figure}
The initial ``slope'' regime can be explained by fluctuations about the naive saddle point in the gravity background (analogous to a quasinormal mode analysis), and is consistent with self-averaging (factorization) so that  $\langle |Z(\beta+it)|^2\rangle_J\approx |\langle Z(\beta+it)\rangle_J|^2$. The ``ramp'' and ``plateau'' are not self-averaging, but are in agreement with the expectations of random matrix theory. The linear ramp results from a $1/(E-E')^2$ term in the correlator of the eigenvalue density $\langle\rho(E)\rho(E')\rangle$, reflecting long-range repulsion
between eigenvalues. The plateau arises from a
modification of this power law when the energy differences are of the order of the typical level spacing $E-E' \sim e^{-S}$. This transition takes place at a time of order
the inverse of this spacing, $t\sim e^{S}$.\footnote{Recently it was argued that this late-time behavior and the non-factorization could be explained by the appearance of wormholes as saddles of the gravitational path integral \cite{Verlinde:2021jwu,Verlinde:2021kgt}. The argument involves the connection between the spectral form factor and the second Renyi entropy of the time
evolved thermofield double state, which involves the so-called ``replica'' wormholes. These geometries arise in the context of the information paradox and are required to obtain a Page curve consistent with unitarity.}

It is important to juxtapose the aforementioned slope-ramp-plateau pattern with systems that exhibit complete integrability. As a simple example, consider the quantum harmonic oscillator, where the energy levels are given by
\be
E_i=\left(i+\frac{1}{2}\right)\omega\,.
\ee
A brief calculation yields,
\be
Z(\beta)=\frac{e^{\beta\omega/2}}{e^{\beta\omega}-1}\,,\qquad g(t)=\frac{1}{2\cosh(\beta\omega)-2\cos(\beta t)}\,,
\ee
unveiling a distinct periodicity in time within the SSF ---see Fig. \ref{fig:SFFH} below.
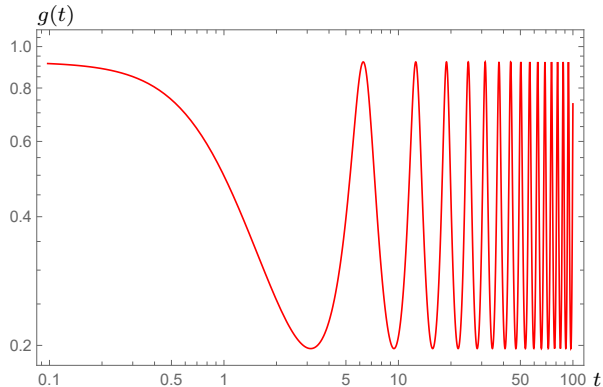
\begin{figure}[h!]
\centering
 \includegraphics[width=3in]{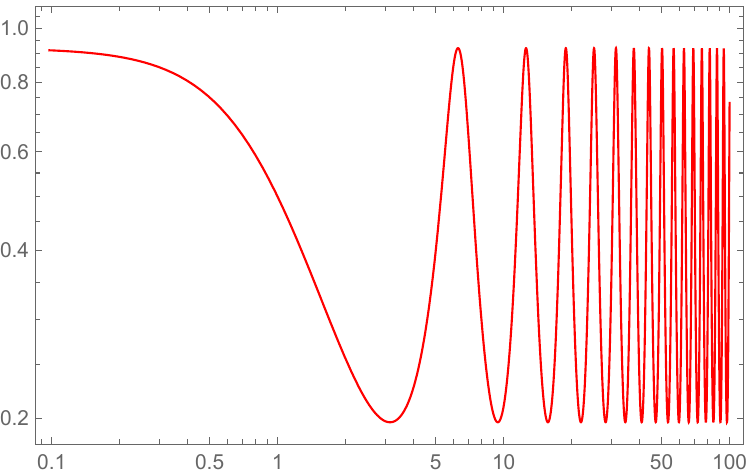}
 \begin{picture}(0,0)
\put(-210,139){\scriptsize $g(t)$}
\put(-2,1){\scriptsize$t$}
\end{picture}
\caption{Typical log-log plot of the spectral form factor in the quantum harmonic oscillator. For the plot we have set parameters $\beta=1$, $\omega=1$.\label{fig:SFFH}}
\end{figure}

It is worth highlighting that unlike the SYK model, our system does not involve disorder or averaging. Instead, it comprises two coupled non-linear oscillators. The questions surrounding disorder averaging, factorization, and wormholes are significant open questions introduced by the SYK model, but may not necessarily be universally applicable to our understanding of black holes. Notably, insights from higher dimensional Conformal Field Theory (CFT) descriptions of black holes offer no indication of these features.

Our overarching aim is to discern which lessons from simple models are genuinely universal in quantum mechanical models with gravity duals. In this context, it is pertinent to note that the distinct slope-ramp-plateau pattern has also been observed in systems with deterministic spectra, obviating the need for disorder or averaging \cite{Das:2023ulz,Das:2023yfj,Das:2023xjr}. For instance, a simple logarithmic spectrum augmented by a noise term exhibits both energy repulsion among eigenvalues and a linear ramp in the SFF, and can be derived from a normal mode analysis of black hole stretched horizons. Consequently, our objective is to ascertain if a similar pattern manifests in the system of two oscillators under our consideration.

Computationally, it is easy to obtain the spectral form factor (\ref{eq:sff}) once we know the spectrum of the system. 
We compute the spectrum numerically, following the methods outlined in Appendix \ref{app:numerics}, and use (\ref{eq:sff}) to estimate the SSF using up to $10^4$ eigenvalues. Higher order eigenvalues will only affect the SSF at very long time scales, provided we work at moderate temperatures. The deviations from our results in this regime are nevertheless uninteresting, as the SSF is dominated there by fluctuations over the final plateau, which approaches the long time average $g(t)\sim Z(2\beta)$. 

\begin{figure}[ht]
  \subfloat[$\mu=0$]{
	\begin{minipage}[c][1\width]{
	   0.3\textwidth}
	   \centering
	   \includegraphics[width=1\textwidth]{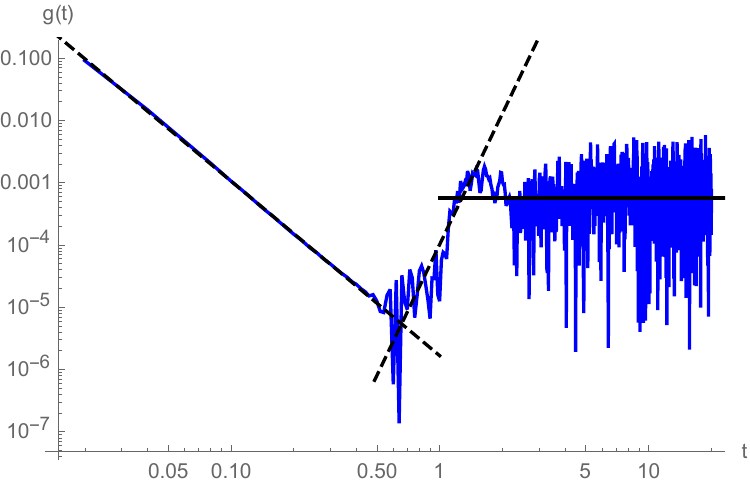}
	\end{minipage}}
 \hfill 	
  \subfloat[$\mu=5$]{
	\begin{minipage}[c][1\width]{
	   0.3\textwidth}
	   \centering
	   \includegraphics[width=1\textwidth]{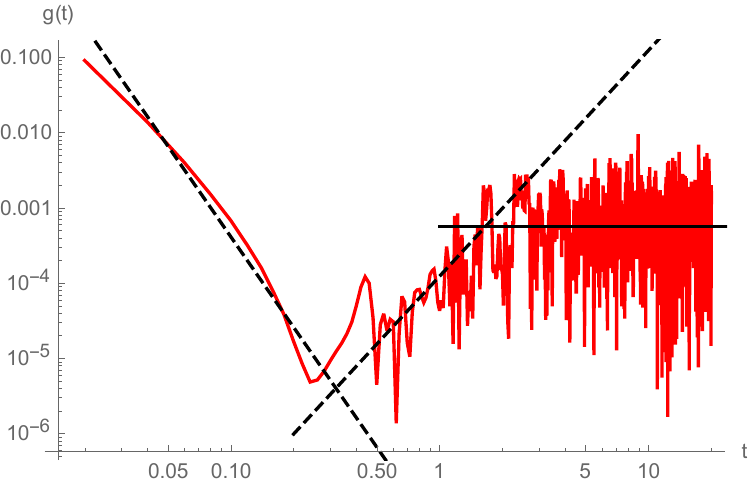}
	\end{minipage}}
 \hfill	
  \subfloat[$\mu=20$]{
	\begin{minipage}[c][1\width]{
	   0.3\textwidth}
	   \centering
	   \includegraphics[width=1\textwidth]{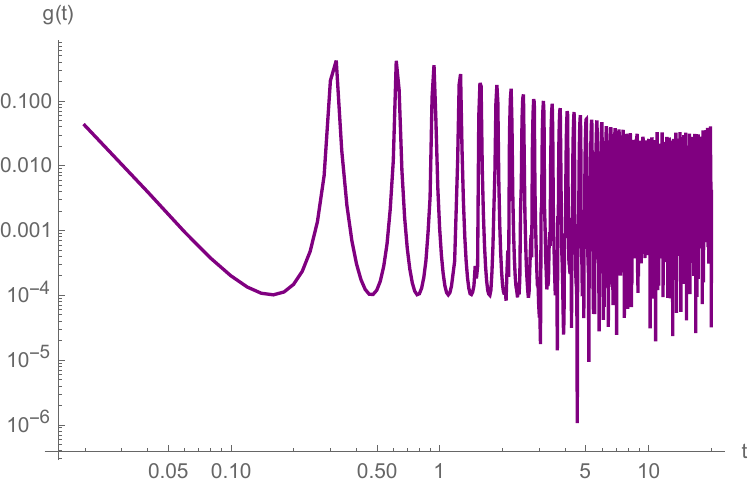}
	\end{minipage}}
\caption{Spectral form factor computed with the even eigenvalues for $\beta=1$ and various values of $\mu$, showing a transition between ``chaotic'' to ``integrable'' behavior.\label{SFFe}}
\end{figure}

\begin{figure}[ht]
  \subfloat[$\mu=0$]{
	\begin{minipage}[c][1\width]{
	   0.3\textwidth}
	   \centering
	   \includegraphics[width=1\textwidth]{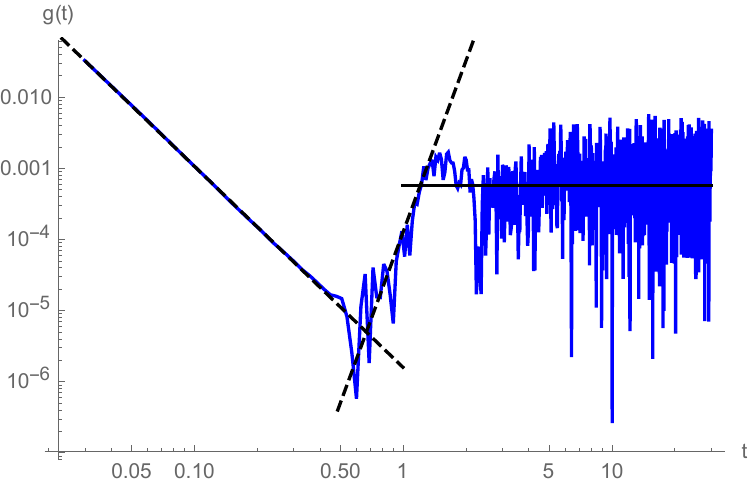}
	\end{minipage}}
 \hfill 	
  \subfloat[$\mu=5$]{
	\begin{minipage}[c][1\width]{
	   0.3\textwidth}
	   \centering
	   \includegraphics[width=1\textwidth]{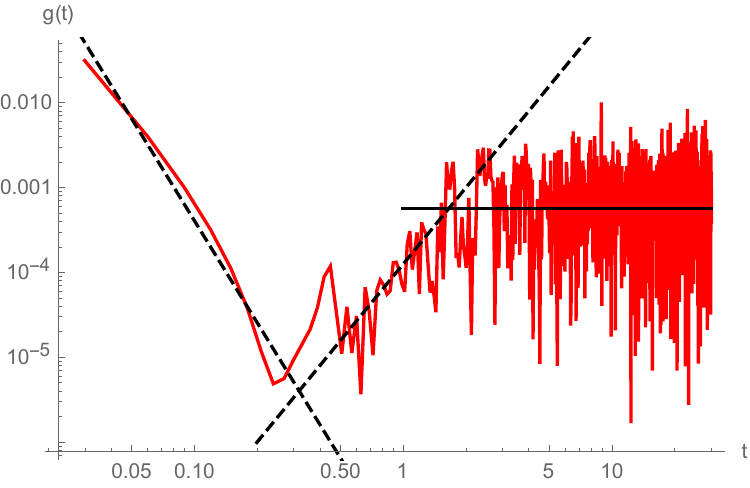}
	\end{minipage}}
 \hfill	
  \subfloat[$\mu=20$]{
	\begin{minipage}[c][1\width]{
	   0.3\textwidth}
	   \centering
	   \includegraphics[width=1\textwidth]{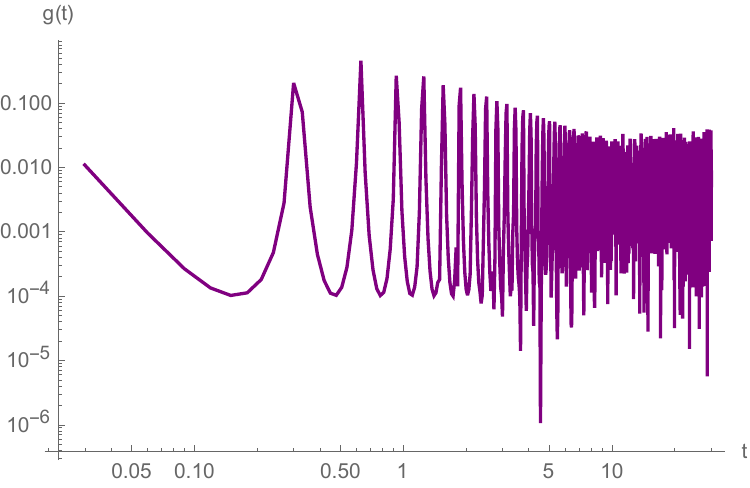}
	\end{minipage}}
\caption{Spectral form factor computed with the odd eigenvalues for for $\beta=1$ and various values of $\mu$, showing a transition between ``chaotic'' to ``integrable'' behavior.\label{SFFo}}
\end{figure}

It is worth noting that the spectrum can be split into two non-interacting sets, even and odd, respectively, given the symmetry $x\to-x$ of the interaction potential. 
In Fig. \ref{SFFe} we plot the SSF obtained in the even sector. 
Similarly, the results for the odd sector are depicted in Fig. \ref{SFFo}. 
In both cases we observe a clear transition between ``chaotic'' to ``integrable'' behavior as we increase the value of the coupling $\mu$. 
However, considering the findings from the previous sections, it is tantalizing to interpret the aforementioned behavior as a consequence of the instability of the system at $\mu\to 0$, rather than a genuine manifestation of quantum chaos.

A couple of final remarks are in order: first, for very small $\mu$ we precisely reproduce the scaling $g(t)\propto 1/t^3$ found in random matrix theory \cite{Cotler:2016fpe}. It is quite remarkable that in a system with only two degrees of freedom we observe such a behavior, though we emphasize this should be attributed to the instability of the system. Second, the slope is generally power law, instead of linear, which indicates some deviations over random matrix theory at intermediate times. We speculate this may be a manifestation of the mixed nature of the phase space for the system under consideration.

\section{Discussion and Outlook}\label{Sec:Conclusions}

In this manuscript, we investigated a system resulting from a truncation of a stringy matrix model conjectured to be dual to a certain gravity theory. We characterized the phase space, finding it to be mixed with an exact integrable point for $\mu=0$ and chaotic behavior for any non-zero $\mu$. When $\mu$ is significantly larger than other scales in the problem, the system simplifies to two decoupled harmonic oscillators and exhibits approximate integrability.

At the quantum level, we observed that the Brody distribution accurately reflects the nature of the classical chaos. The parameter $b$, characterizing the eigenvalue distribution as an interpolation between Poisson and Wigner, nicely dovetails the classical results. Specifically, very small and very large values of $\mu$ yield Poisson-like distributions, whereas intermediate values of $\mu$ align more closely with Wigner distributions.  

We have also accurately identified a puzzling situation: {\it the $\mu=0$ system is classically integrable, yet some of the indicators of quantum chaos have turned out positive.} Let us, therefore, discuss the two quantum chaos indicators whose results disagree with the classical analysis of chaos:  the quantum Lyapunov exponent as computed from the OTOC and the spectral form factor.

There have been various studies on the behavior of the OTOC demonstrating exponential growth in systems that are not classically chaotic. For instance, in \cite{Hashimoto:2020xfr}, the authors examined the inverted harmonic oscillator and observed exponential growth of the OTOC. Our model aligns well with this paradigm, particularly for $\mu=0$, where the presence of flat directions renders the system unstable. Other instances of exponential growth in OTOCs due to instabilities in integrable systems have been documented in several contexts \cite{Pappalardi:2018frz,Hummel:2018brt,Pilatowsky-Cameo:2019qxt,Xu:2019lhc,Chavez-Carlos:2022qqj}.  A tantalizing modification for the OTOC  in mixed system was put forward in \cite{Trunin:2023xmw,Trunin:2023rwm,Kolganov:2022mpe}.

To gain deeper insights into the nature of quantum chaos in our system, we conducted a study of the late-time behavior of the OTOC. We observed that for $\mu=0$, the system exhibits more irregular behavior compared to higher values of $\mu$, where regularity increases. Regarding our analysis using the spectral form factor, we identified a clear transition from chaotic behavior at $\mu=0$ to integrable behavior at large $\mu$. These findings align with previous studies, such as those on the two-body Sachdev-Ye-Kitaev model, which also show a transition in the spectral form factor from an exponential ramp in integrable systems to a linear ramp in chaotic systems \cite{Winer:2020mdc}.

Our collective results can be interpreted within a coherent framework consisting of three distinct regions:
\begin{enumerate}
\item At $\mu=0$, the system is integrable but \emph{unstable}, leading to an exponential growth of the OTOC and a distinct slope-ramp-plateau pattern in the SFF without exhibiting chaos or scrambling behavior.
\item
For non-zero but moderately small values of $\mu$, the system exhibits classical and quantum chaos.
\item At very large values of $\mu$, the system approaches integrability at the classical level and becomes stable, resulting in a decrease of all quantum chaos indicators.
\end{enumerate}
This breakdown provides a clear understanding of how the behavior of the system evolves with respect to the parameter $\mu$.

\section*{Future directions}
Let us conclude by highlighting some natural future directions. The ultimate goal is to find quantum chaotic behavior, compatible with a gravitational interpretation, in simple quantum mechanical systems. There are three main directions may be worth exploring: 

\begin{itemize}

\item Our truncation to only two effective degrees of freedom may be too drastic to capture gravity-like behavior, which is typically associated with theories possessing a large number of degrees of freedom. We hope that by retaining increasingly more degrees of freedom, gravity-like behavior could emerge. An intriguing question concerns the minimal number of degrees of freedom necessary to start detecting gravity-like phenomena such as, for example, the saturation of the bound on chaos. We expect to report on these efforts elsewhere. An optimistic outlook stems from recent studies, such as the investigation of $SU(N=3)$ in the BFSS matrix model \cite{Hanada:2009ne}, which reported finding conformal behavior in correlators. Recall that in the large-$N$ limit, the gauge/gravity correspondence predicts a particular scaling behavior in the IR region for the two-point correlators of operators corresponding to certain supergravity modes. More recently, the necessary number of qubits required to uncover certain black hole features was estimated in \cite{Maldacena:2023acv} to be about $7000$, corresponding to  $N=16$.

\item Our setup is equipped with the necessary ingredients to compute von Neumann entropies. For example, we can trace over one of the two oscillators to generate a reduced density matrix. Under suitable conditions, this setup could reveal a form of a Page curve, mirroring quantum information phenomena in black hole evaporation. More broadly, exploring information-theoretic quantities in stringy matrix models more comprehensively could yield valuable insights. Specifically, studying the BFSS model ($\mu=0$) could facilitate direct comparisons with gravitational data, as demonstrated in previous works \cite{Hanada:2013rga,Hanada:2008ez}. These investigations hold promise for deeper understanding at the intersection of quantum mechanics and gravity. 

\item Lastly, we could analyze other indicators of quantum chaos, to further characterize mixed phase spaces. A promising avenue is quantum complexity, particularly Krylov-based measures of complexity \cite{Parker:2018yvk,Balasubramanian:2022tpr,Caputa:2024vrn}. Historically, complexity was introduced in the context of black holes to address the puzzle of why black hole interiors continue to grow after scrambling \cite{Susskind:2014moa}. Since then, numerous proposals for holographic complexity have been suggested, with `complexity=volume' and `complexity=action' being the most prominent ones \cite{Stanford:2014jda,Brown:2015bva}. While a direct mapping between these proposals and boundary notions of complexity remains elusive, recent work has linked Krylov state complexity to `complexity=volume' \cite{Rabinovici:2023yex}, within the context of 2-dimensional JT gravity and the double-scaled SYK model. Moreover, Krylov-based complexity measures have been effectively employed in quantum mechanical models to discern between integrable and chaotic behaviors \cite{Bhattacharjee:2022vlt,Rabinovici:2022beu,Erdmenger:2023wjg,Camargo:2023eev,Huh:2023jxt,Camargo:2024deu} (see \cite{Nandy:2024htc} for a review).

\end{itemize}

\section*{Acknowledgements}
It is a pleasure to thank Marcos Rigol for useful discussions and Shahin Sheikh-Jabbari for comments on the manuscript. LPZ is partially supported by the U.S. Department of Energy under grant DE-SC0007859, he also acknowledges support from an IBM Einstein Fellowship while at the Institute for Advanced Study in 2022/2023. JFP is supported by the `Atracci\'on de Talento' program grant 2020-T1/TIC-20495 and by the Spanish Research Agency through the grants CEX2020-001007-S and PID2021-123017NB-I00, funded by MCIN/AEI/10.13039/501100011033 and by ERDF A way of making Europe. The work of NQ was supported by a Sistema Nacional de Investigadores grant and PROSNI-2020/23 from Universidad de Guadalajara.


\appendix
\section{Numerical Methods\label{app:numerics}}


Here we provide some details pertaining to our numerical methods. We conveniently separate the Hamiltonian into an unperturbed free  part and a perturbation, by writing it as
\begin{eqnarray}
\hat{H} &=& \hat{H}_0 + \tilde{V}(x,y), \nonumber \\
\hat{H}_0 &\equiv & - \frac{1}{2} \frac{d^2}{dx^2} - \frac{1}{2} \frac{d^2}{dy^2}  
+ \frac{\alpha ^4 x^2}{2}+\frac{\beta ^4 y^2}{2} \nonumber \\
\tilde{V}(x,y) &\equiv &  \frac{x^4}{2}+\frac{\mu^2 x^2}{8}+x^2 y^2+\frac{y^4}{2}-\mu  y^3+\frac{\mu ^2 y^2}{2} - \frac{\alpha ^4 x^2}{2} - \frac{\beta ^4 y^2}{2}.
\end{eqnarray}
where  $\alpha$ and $\beta$ are arbitrary (real) parameters. As a consequence of this splitting, $\hat{H}_0$ is the Hamiltonian of a two--dimensional harmonic oscillator. Clearly, the eigenvalues and eigenfunctions for $\hat{H}_0$ are 
\begin{eqnarray}
    \hat{H}_0  \Psi_{n,m}(x,y) &=& E_{nm} \Psi_{n,m}(x,y), \qquad E_{nm} \equiv  \left[\left( n +  \frac{1}{2 }\right) \alpha^2 +  \left( m +  \frac{1}{2 }\right) \beta^2  \right], \nonumber \\
\Psi_{n,m}(x,y) &=&  \psi_{n}(\alpha, x)  \psi_{m}(\beta,y), \label{eq_basis}  \qquad 
\psi_{n}(\alpha, x) = e^{-\frac{1}{2} \alpha^2 x^2} \sqrt{\frac{2^{-n} \ \alpha  }{\sqrt{\pi }n!}} \ H_n(x \alpha ).
\end{eqnarray}

In order to solve accurately the  Schr\"odinger equation for the full Hamiltonian
\begin{equation}
\hat{H}  \Phi_{n,m}(x,y) = {\cal E}_{n,m} \Phi_{n,m}(x,y),
\label{eq_Sch}
\end{equation}
we may expand the eigenfunctions $\Phi_{n,m}(x,y)$ in the basis of eigenfunctions of $\hat{H}_0$
\begin{equation}
\Phi_{n,m}(x,y) = \sum_{n,m} c_{n,m} \Psi_{n,m}(x,y),
\label{eq_decomp}
\end{equation}
where both terms in the r.h.s. of the equation depend on the parameters $\alpha$ and $\beta$ 

{\it Key numerical improvement:} Although this expansion is valid for any (real) value of the parameters, $(\alpha, \beta)$, it is desirable to select values for which the convergence of the expansion is faster: in this case a given precision 
can be achieved with a smaller number of terms (see \cite{amore2005new,Amore_2006,Amore_2007} for more details on this method). . 

Before describing the procedure that we have adopted to select the parameters, it is worth noting that $\tilde{V}(x,y)$ is even with respect to $x \rightarrow -x$ and therefore one can split the basis into two sets
\begin{equation}
\Psi^{even}_{nm}(x,y)  =  \Psi_{2n,m}(x,y) , \qquad 
\Psi^{odd}_{nm}(x,y)   =  \Psi_{2n+1,m}(x,y) 
\end{equation}

In this way, we can calculate separately the even and odd (in $x$) eigenfunctions of eq.~(\ref{eq_Sch}). By using the appropriate decomposition in eq.~(\ref{eq_decomp}), and working with a finite number of
element of the basis of $\hat{H}_0$ (either even or odd in $x$), we obtain a matrix representation
for $\hat{H}$.  The matrix element of $\hat{H}$ are obtained explicitly in terms of the matrix elements $\langle n | x^{2r} | l \rangle$ and $\langle n | x^{2r+1} | l \rangle $, $r=0,1,2,\ldots$. Notice that the initial Hamiltonian (and therefore both its eigenvalues and eigenfunctions) is independent of the parameters, $\alpha$ and $\beta$, but the eigenvalues and eigenfunctions obtained using any finite--dimensional  representation will necessarily have a spurious dependence on the parameters.
For this reason, a careful identification of the optimal values of $\alpha$ and $\beta$ is needed in order to suppress this artificial dependence and thus obtain more precise numerical results, while working with a finite number of elements. 

A crucial observation is that the trace of $\hat{H}$, which is invariant under unitary transformations, should be independent of the basis (and therefore of the parameters specifying the basis). However, if one works with a finite number of states, the trace displays an artificial dependence of the parameters. 
In previous works, refs.~\cite{amore2005new,Amore_2006,Amore_2007}, it has been observed that one can minimize this dependence by selecting  the parameters that are solutions of the equations
\begin{equation}
\frac{d T_N}{d\alpha} = 0, \qquad  \hspace{1cm}\frac{d T_N}{d\beta} = 0,
\label{eq_PMSa}
\end{equation}
where $T_N$ is the trace of the Hamiltonian in the truncation defined by $N$. This procedure allows one to work with a nearly optimal basis, with a limited computational cost (the identification of the optimal parameters does not require the diagonalization of the matrix). 

In the bulk of the paper we consider large sets of the form  $N_x=300$ and $N_y=300$, leading to sparse matrices (even and odd) of size approximately $45000 \times 45000$.


\bibliographystyle{JHEP}
\bibliography{refs-CQM}

\end{document}